\documentclass[twoside]{article}
\usepackage{float}
\usepackage{amssymb}
\usepackage{amstext,enumerate}
\usepackage{bm}
\usepackage{graphics}
\usepackage{graphicx}
\usepackage{amsthm}
\usepackage{amsmath}
\usepackage{latexsym}
\usepackage{rotate}
\usepackage{blkarray}
\usepackage{lscape}
\usepackage{color}
\usepackage{epsfig,epsf,psfrag}
\usepackage{sectsty}
\usepackage{graphics}
\usepackage{epsfig}
\usepackage{multirow}
\usepackage{graphicx}
\usepackage{amsmath}
\usepackage{url}
\usepackage{hyperref}
\usepackage{booktabs}
\usepackage{amsthm}
\usepackage{sectsty}
\usepackage{epsfig}
\usepackage{rotate}
\usepackage{subfigure}
\usepackage{lscape}
\usepackage{setspace}
\usepackage{subcaption}
\usepackage{float}
\usepackage{mathtools}
\usepackage{algorithmic}
\usepackage{algorithm}

\usepackage{siunitx}
\newcommand{\se}[1]{\multicolumn{1}{S[table-format=2.3,parse-numbers=false]}{(#1)}}

\usepackage[accepted]{aistats2026}
\usepackage[round]{natbib}

\newcommand{\F}{Fr\'echet }
\newcommand{\real}{{\mathbb{R}}}
\newcommand{\M}{{\mathcal M}}
\newcommand{\up}{{\upsilon}}
\newcommand{\bX}{{\textbf X}}
\newcommand{\bx}{{\textbf x}}
\DeclareMathOperator*{\argmin}{arg\,min}

\begin{document}

%

%

\twocolumn[

\aistatstitle{ADOPT: Additive Optimal Transport Regression}

\aistatsauthor{ Wookyeong Song \And Hans-Georg M\"uller }

\aistatsaddress{ University of California, Davis \And University of California, Davis } ]

\begin{abstract}
Regression models for responses $Y$ taking values in general metric spaces $(\mathcal{M}, d)$, with Euclidean predictors $X \in \mathbb{R}^p,$ has attracted growing interest in recent years.
  While additive regression is a powerful tool for enhancing interpretability and mitigating the curse of dimensionality in the presence of multivariate predictors, its direct extension is hindered by the absence of vector space operations in general metric spaces. We propose a novel framework for additive optimal transport regression, which incorporates additive structure through optimal geodesic transports. A key idea is to extend the notion of optimal transports in Wasserstein spaces to general geodesic metric spaces. This unified approach accommodates a wide range of responses, including probability distributions, symmetric positive definite (SPD) matrices with various metrics and spherical data. The practical utility of the method is illustrated with correlation matrices derived from resting state fMRI brain imaging data.
\end{abstract}

\section{INTRODUCTION}

We study regression where the response is a random object $Y$ in a geodesic space $(\M, d)$ with general metric $d$ and the predictor is Euclidean $X \in \real^p.$ Such settings arise widely for  modern non-Euclidean data \citep{petersen2019frechet, schotz2022nonparametric}, including diffusion tensor imaging in the space of symmetric positive definite matrices \citep{lin2023additive}, mortality analysis in the space of one-dimensional  probability distributions with the Wasserstein geometry \citep{mull:23:2},  functional brain networks data \citep{fornito2016fundamentals, zhou2022network}, corpus linguistics data on Riemannian manifolds \citep{severn2022manifold}, functional data \citep{wang2016functional} and many others \citep{marron2021object,song2026non}.

Local \F regression extends kernel-based local linear regression to metric-valued responses 
\citep{petersen2019frechet, chen2022uniform, schotz2022nonparametric}. However, like standard nonparametric regression, it suffers from the curse of dimensionality when $p>1$, 
which limits its practical use for multivariate predictors.  Dimension reduction via single-index models has been proposed to address this issue \citep{bhattacharjee2025nonlinear, bhattacharjee2023single, ghosal2023frechet, hong2025robust, zhang2024dimension}, but compressing predictor information into a single index may be too restrictive in some cases and then lead to biases.  For the case of Euclidean responses,  additive models have been shown to provide a competitive  alternative that mitigates the curse of dimensionality while offering interpretability and flexibility \citep{linton1995kernel, yu2008smooth}.

The generalization of additive models for metric space-valued responses is challenging because vector space operations, such as addition or scalar multiplication are unavailable in general metric spaces.  Existing work covers additive structures for non-Euclidean responses only in some specialized settings, such as Hilbert space-valued responses \citep{jeon2020additive}, distributional responses via extrinsic approaches \citep{han2020additive} and Lie group–valued data \citep{lin2023additive}. We introduce a novel additive framework based on geodesic optimal transports that generalize  optimal transports in Wasserstein spaces to scenarios that feature  general geodesic metric spaces. To estimate the additive transport map, we develop a transport backfitting algorithm, which is an extension of the popular classical backfitting method \citep{breiman1985estimating, hastie1986generalized}. 

\textbf{Outline of the paper}: In Section 2 and 3 we review related works and preliminaries regarding   \F means, local \F regression and geodesic optimal transport. Section 4 presents the proposed additive optimal transport regression (ADOPT) model and associated  estimation procedure using a transport backfitting algorithm and theory  on oracle convergence. Section 5 reports numerical experiments and Section 6  real data illustrations for fMRI correlation matrices. 

\section{RELATED WORK}

Prevalence of non-Euclidean data in recent years has created demand for principled and interpretable statistical tools for their analysis. A promising approach is to  treat the sample elements as points in a metric space $(\M, d)$ \citep{dubey2024metric, wang2024nonparametric}. Unlike Euclidean settings, general metric spaces lack algebraic structure, motivating  distance-based statistical methods. 
The selected metric can greatly affect statistically relevant properties of the metric space \citep{song2025inference}. For example for the case of symmetric positive definite (SPD) matrices the log-Cholesky metric \citep{lin2019riemannian} has desirable  properties such as avoiding the swelling effect and allowing for closed-form parallel transport along geodesics. 

The regression analysis of metric-valued responses has been studied in specialized  scenarios  \citep{yuan:12, corn:17}. A general approach is \F regression \citep{petersen2019frechet}, based on the concept of conditional \F means.  \F regression makes use of the vector space structure of predictors in $\mathbb{R}^p$ to find elements in the metric space that minimize a weighted squared distances, and various extensions have been developed \citep{song2023errors}.

For distributional responses with the Wasserstein metric, geodesics are represented by optimal transport maps, enabling rich statistical applications \citep{chewi2024statistical}. Transport-based regression models have been developed for distribution-on-distribution regression \citep{ghodrati2022distribution} and autoregressive models  \citep{zhu2023autoregressive}. { \citet{zhu2025geodesic} proposed a geodesic optimal transport (GOT) regression model, extending a parametric multiple regression framework via transport maps. However, this model treats the transport maps as fixed covariates and restricts estimation to finite-dimensional multiplicative parameters in $\mathbb{R}^p$.  However nonparametric additive extensions that directly estimate transport maps considering  the intrinsic geometry of the space have remained unexplored. We address this gap by proposing an additive optimal transport regression.}

\section{PRELIMINARIES}

\subsection{\F mean and \F regression}

Let $(\mathcal{M}, d)$ be a metric space with distance $d: \mathcal{M} \times \mathcal{M} \to \mathbb{R}$. Consider random objects $(\mathbf{X}, Y) \sim \mathcal{F}$, where  $\mathcal{F}$ is their joint distribution, $\mathbf{X} = (X_{1}, \ldots, X_{p}) \in \prod_{j=1}^{p}\mathcal{X}_{j}$ with compact domain $\mathcal{X}_{j} \subset \mathbb{R}$, and $Y \in \mathcal{M}.$

The \F mean \citep{frec:48} generalizes the Euclidean expectation to random objects $Y$ situated in metric spaces, with population and sample versions 
\begin{align}
        \label{eq:fm} \mathbb{E}_{\oplus}Y &= \argmin_{\up \in \mathcal{M}} \mathbb{E}{d^2(Y,\up)},    \\
        \label{eq:fm_samp}\hat{\mathbb{E}}_{\oplus}Y &=  \argmin_{\up\in\mathcal{M}} \frac{1}{n} \sum_{i=1}^{n}d^2(Y_i, \up).
\end{align}
Since $\mathcal{M}$ lacks vector space structure, Euclidean regression models cannot be directly applied. The \F regression function is  an implementation of conditional \F means, extending conditional expectation to metric-valued responses:
\[\mathbb{E}_{\oplus} [Y \mid \bX = \bx] = \argmin_{\up\in\mathcal{M}}\mathbb{E}\{d^2(Y, \up)\mid \bX=\bx\}.\] For scalar responses $Y \in \real,$  \F conditional means  reduce to the standard conditional expectation. 

Local \F regression \citep{petersen2019frechet} adapts local linear regression by replacing Euclidean distances with $d$ and using kernel-based weights. We consider the univariate predictor $X \in \mathbb{R}$ with kernel $K$, bandwidth $h>0$, and rescaled kernel $K_h(\cdot)=h^{-1}K(\cdot/h)$, where the extension to $X\in\mathbb{R}^p$ with $p>1$ is straightforward. This leads to 
\[\mathbb{E}_{\oplus L}[Y \mid X = x]=\argmin_{\up\in\mathcal{M}}\mathbb{E}\{w(x, h)d^2(Y, \up)\},\]
with  $w(x, h)=K_h(X-x)\{u_2-u_1(X-x)\}/\sigma_0^2$,  $u_j=\mathbb{E}\{K_h(X-x)(X-x)^j\}$, $j=0, 1, 2$  and $\sigma_0^2=u_0u_2-u_1^2$. 

Given random samples $(X_i, Y_i) \in \real \times \M,$ $i = 1, 2, \ldots, n,$ the corresponding sample estimator is 
\begin{align}
\begin{split}
    &\hat{\mathbb{E}}_{\oplus L}[Y \mid X = x]=\argmin_{\up\in\mathcal{M}}\frac{1}{n}\sum_{i=1}^n\hat{w}_{i}(x, h)d^2(Y_i, \up),\\
    &\text{where }\hat{w}_{i}(x, h)=K_h(X_i-x)\{\hat{u}_2-\hat{u}_1(X_i-x)\}/\hat{\sigma}_0^2,
    \label{eq:loc_freg_samp}
\end{split}
\end{align} with $\hat{u}_j=n^{-1}\sum_{i=1}^nK_h(X_i-x)(X_i-x)^j$ for $j=0, 1, 2$ and $\hat{\sigma}_0^2=\hat{u}_0\hat{u}_2-\hat{u}_1^2$. For various asymptotic results on the  convergence of this estimator see  \citet{petersen2019frechet, chen2022uniform}.

\subsection{Geodesic optimal transport}
For a metric space $(\mathcal{M},d)$, the length of a path $\gamma: \left[0, 1 \right] \rightarrow \mathcal{M}$ is defined as 
\begin{align*}
    l(\gamma) = \sup_{0=t_{0}<t_{1}<\ldots<t_{n}=T, n \in \mathbb{N}} \sum_{i=1}^{n}d(\gamma\left(t_{i-1}),\gamma(t_{i})\right).
\end{align*} A path $\gamma_{\upsilon_{1}, \upsilon_{2}}$ is a geodesic from $\upsilon_{1}$ to $\upsilon_{2} $ if $\gamma_{\upsilon_{1}, \upsilon_{2}}(0) = \upsilon_{1},$ $\gamma_{\upsilon_{1}, \upsilon_{2}}(1) = \upsilon_{2}$ and $d\left(\gamma_{\upsilon_{1}, \upsilon_{2}}(t),\gamma_{\upsilon_{1}, \upsilon_{2}}(t')\right) \propto \lvert t-t' \rvert$ for all $t, t' \in \left[0, 1 \right]$. A metric space $(\mathcal{M},d)$ is called a geodesic space if every pair of points $\upsilon_{1}, \upsilon_{2} \in \mathcal{M}$ can be joined by a geodesic, and a  unique geodesic space if this geodesic is unique \citep{burago2022course, lang2012differential}. In Euclidean space, geodesics are straight lines that connect two points.  Throughout we assume that $(\mathcal{M},d)$ is a bounded, separable, and uniquely geodesic space. For geodesic optimal transports, we require the following  
ubiquity assumption \citep{zhu2025geodesic}:  

\noindent \textbf{Assumption 1} For any random objects $\up_{1}, \up_{2}, \up_{3}  \in \mathcal{M},$ where $\up_{1} \neq \up_{2},$ there exists a map $\Gamma : \mathcal{M} \times \mathcal{M} \times \mathcal{M} \rightarrow \mathcal{M}$ and a unique element $\up_{4} \in \mathcal{M}$ such that
\begin{align*}
    \Gamma(\up_{1}, \gamma_{\up_{1}, \up_{2}}(r), \up_{3}) = \gamma_{\up_{3},\up_{4}}(r),
\end{align*}
for any $r \in \left[0,1\right].$ 

{This ubiquity assumption ensures that for any geodesic $\gamma_{\up_{1}, \up_{2}}$ and any point $\up_{3}  \in \mathcal{M},$ there exists a unique geodesic starting from $\up_{3}$ whose direction matches that of $\gamma_{\up_{1}, \up_{2}}.$ In other words, any geodesic segment can be attached at any point in $\M,$ producing a new geodesic with endpoint $\up_4 \in \M.$} In Euclidean space $\real^p,$ the map is given by $\Gamma(\up_{1}, \gamma_{\up_{1}, \up_{2}}(r), \up_{3}) = \up_3 + r(\up_2 - \up_1),$ for $r \in \left[0,1\right],$ and $\up_4 = \up_3 + (\up_2 - \up_1).$ 

\noindent \textbf{Definition 1 (Geodesic optimal transport)} \textit{Under Assumption 1, for any random objects $\up_{1}, \up_{2} \in \mathcal{M},$ the geodesic transport and inverse geodesic transport from $\up_{1}$ to $\up_{2},$ denoted as $T_{\up_{1},\up_{2}} : \mathcal{M} \rightarrow \mathcal{M},$ $ T^{-1}_{\up_{1},\up_{2}} : \mathcal{M} \rightarrow \mathcal{M},$ respectively, are defined as
\begin{align*}
\begin{split}
    T_{\up_{1},\up_{2}}(\up) = \Gamma(\up_{1}, \up_{2}, \up),~
    T^{-1}_{\up_{1},\up_{2}}(\up) =  \Gamma(\up_{2}, \up_{1}, \up), 
\end{split} 
\end{align*} for all  $\up \in \mathcal{M}.$
Furthermore, the set of all geodesic transports maps is 
    $\mathcal{T} = \left\{T_{\up_{1},\up_{2}} : \up_{1}, \up_{2} \in \mathcal{M} \right\}.$}

We introduce an addition operation in the transport space $\mathcal{T}$ as function composition:
\begin{align*}
    T_{\up_{1},\up_{2}} \oplus T_{\up_{3},\up_{4}} = T_{\up_{3},\up_{4}} \circ T_{\up_{1},\up_{2}},
\end{align*}
for any $T_{\up_{1},\up_{2}}, T_{\up_{3},\up_{4}} \in \mathcal{T}$ \citep{zhu2023autoregressive}. {For intuition, consider the Euclidean case $\M = \mathbb{R}^p.$ For any $\up \in \mathbb{R}^p,$  \begin{align*}
    &\left[ T_{\up_{1},\up_{2}} \oplus T_{\up_{3},\up_{4}} \right](\up) = T_{\up_{3},\up_{4}} (T_{\up_{1},\up_{2}} (\up)) \\
    &~= T_{\up_{3},\up_{4}} \left(\up + (\up_2 - \up_1)\right) = \up + (\up_2 - \up_1) + (\up_4 - \up_3),
\end{align*} so composing transports with different start and end points simply adds the corresponding displacement vectors. An analogous principle holds in Hilbert spaces and Riemannian manifolds $\M$ through parallel transport along geodesics.} 

\noindent \textbf{Example 1 (Space of distributional data with the Wasserstein metric)} 

Let $(\mathcal{W}_{2}(\mathcal{D}), d)$ be the space of one-dimensional probability distributions on a compact domain $\mathcal{D} \subset \real$ with finite second moments with $2$-Wasserstein metric $d,$ \begin{align*}
    d^2(\up_1, \up_2) = \int_{0}^{1} \left\{F^{-1}_1(s) - F^{-1}_2(s) \right\}^2 ds,
\end{align*} where $F^{-1}_1(\cdot)$ and $F^{-1}_2(\cdot)$ are the quantile functions of $\up_1$ and $\up_2,$ respectively. 

Let $\up \in \mathcal{W}_{2}(\mathcal{D})$ be a random element of $(\mathcal{W}_{2}(\mathcal{D}), d)$
For any measurable function $l : \mathcal{D} \rightarrow \mathcal{D}$ and $\up \in \mathcal{W}_{2}(\mathcal{D}),$ let $l \# \up$ denote the push-forward measure of $\up,$ satisfying $l \# \up(A) = \up\left(\left\{x : l(x) \in A  \right\} \right)$ for all $A \in \mathcal{B}(\mathcal{D}),$ the Borel $\sigma$-algebra on $\mathcal{D}$.   For any two distributions $\up_{1}, \up_{2} \in \mathcal{W}_{2}(\mathcal{D}),$ the optimal transport map is 
\begin{align*}
    OT_{\up_{1},\up_{2}} = F^{-1}_{2} \circ F_{1}.
\end{align*}
The geodesic $\gamma_{\up_{1}, \up_{2}}$ is given by McCann's interpolant \citep{mccann1997convexity},
\begin{align*}
    \gamma_{\up_{1}, \up_{2}}(r) = (\mathrm{id} + r(OT_{\up_{1},\up_{2}} - \mathrm{id})) \# \up_{1}, \quad r \in \left[0,1\right],
\end{align*}
where $\mathrm{id}: \mathcal{D} \rightarrow \mathcal{D}$ is the identity map. For notational convenience, we simply write $\mathcal{W}_2(\mathcal{D})$  as $\mathcal{W}_2.$

For any random objects, $\up_{1}, \up_{2}, \up_{3} \in \mathcal{W}_{2},$ where $\up_{1} \neq \up_{2},$  assumption 1 is satisfied with
\begin{align*}
\begin{split}
        \Gamma(\up_{1}, \gamma_{\up_{1}, \up_{2}}(r), \up_{3}) &= (\mathrm{id} + r(OT_{\up_{1},\up_{2}} - \mathrm{id})) \# \up_{3} \\
        &= \gamma_{\up_{3}, \up_{4}}(r),\quad r \in \left[0,1\right], 
\end{split}
\end{align*}
where $\up_{4} = T_{\up_{1}, \up_{2}}(\up_{3}) = OT_{\up_{1},\up_{2}} \# \up_{3}.$ \\~


\noindent \textbf{Example 2 (SPD matrices with the log-Cholesky metric)} 

Let $\mathcal{S}^{+}_{m}$ be the collection of $m \times m$ SPD matrics. For any $S \in \mathcal{S}^{+}_{m},$ there exists a unique lower triangular matrix $L$ from the Cholesky decomposition, such that $S = LL^{\mathrm{T}},$ where $L$ is composed of  $\lfloor L \rfloor,$ the strictly lower triangular part, and $D(L),$ the diagonal part. The log-Cholesky metric is defined as \begin{align*}
    &d^2(S_{1}, S_{2}) \\
    &= \lVert \lfloor L_{1} \rfloor - \lfloor L_{2} \rfloor \rVert^{2} + \lVert \log(D(L_{1})) - \log(D(L_{2})) \rVert^{2},
\end{align*} where $S_1 = L_1L_1^{\mathrm{T}},$ $S_2 = L_2L_2^{\mathrm{T}},$ and $\lVert \cdot \rVert$ is the Frobenius norm. 
The geodesic $\gamma_{S_{1}, S_{2}}$ is given by
\begin{align*}\begin{split} 
    &\gamma_{S_{1}, S_{2}}(r) = L(r) L(r)^{\mathrm{T}},\\ 
        &L(r) = \lfloor L_{1} \rfloor + r( \lfloor L_{2} \rfloor - \lfloor L_{1} \rfloor)\\
        &\quad + \exp\left[ \log(D(L_{1})) + r (\log(D(L_{2})) - \log(D(L_{1})))   \right],
    \end{split}
\end{align*} for any $r \in \left[0, 1 \right].$ For any random objects, $S_{1}, S_{2}, S_{3} \in \mathcal{S}^{+}_{m},$ where $S_{1} \neq S_{2}$ and $S_3 = L_3 L_3^{\mathrm{T}},$ Assumption 1 is satisfied for all $r \in \left[0, 1 \right]$  with
\begin{align*}
    \begin{split}
        &\Gamma(S_{1}, \gamma_{S_{1}, S_{2}}(r), S_{3}) = L_\Gamma(r)L_\Gamma(r)^{\mathrm{T}} = \gamma_{S_{3},S_{4}}(r),\\
           &L_\Gamma(r) = \lfloor L_{3} \rfloor + r( \lfloor L_{2} \rfloor - \lfloor L_{1} \rfloor)\\
           &\quad + \exp\left[ \log(D(L_{3})) + r (\log(D(L_{2})) - \log(D(L_{1})))   \right],
    \end{split}
\end{align*} $S_3 = L_3 L_3^{\mathrm{T}}, \, S_4 = L_\Gamma(1) L_\Gamma(1)^{\mathrm{T}}.$

Geodesic transports satisfying Assumption 1 can be constructed for several metrics on $\mathcal{S}_{m}^{+}$ including the Frobenius and power Frobenius metric and also for the space of graph Laplacians representing networks \citep{zhu2025geodesic}.


\section{ADDITIVE OPTIMAL TRANSPORT REGRESSION}

\subsection{Population Model}
\begin{figure*}[ht]
    \centering
    \includegraphics[width=0.45\linewidth]{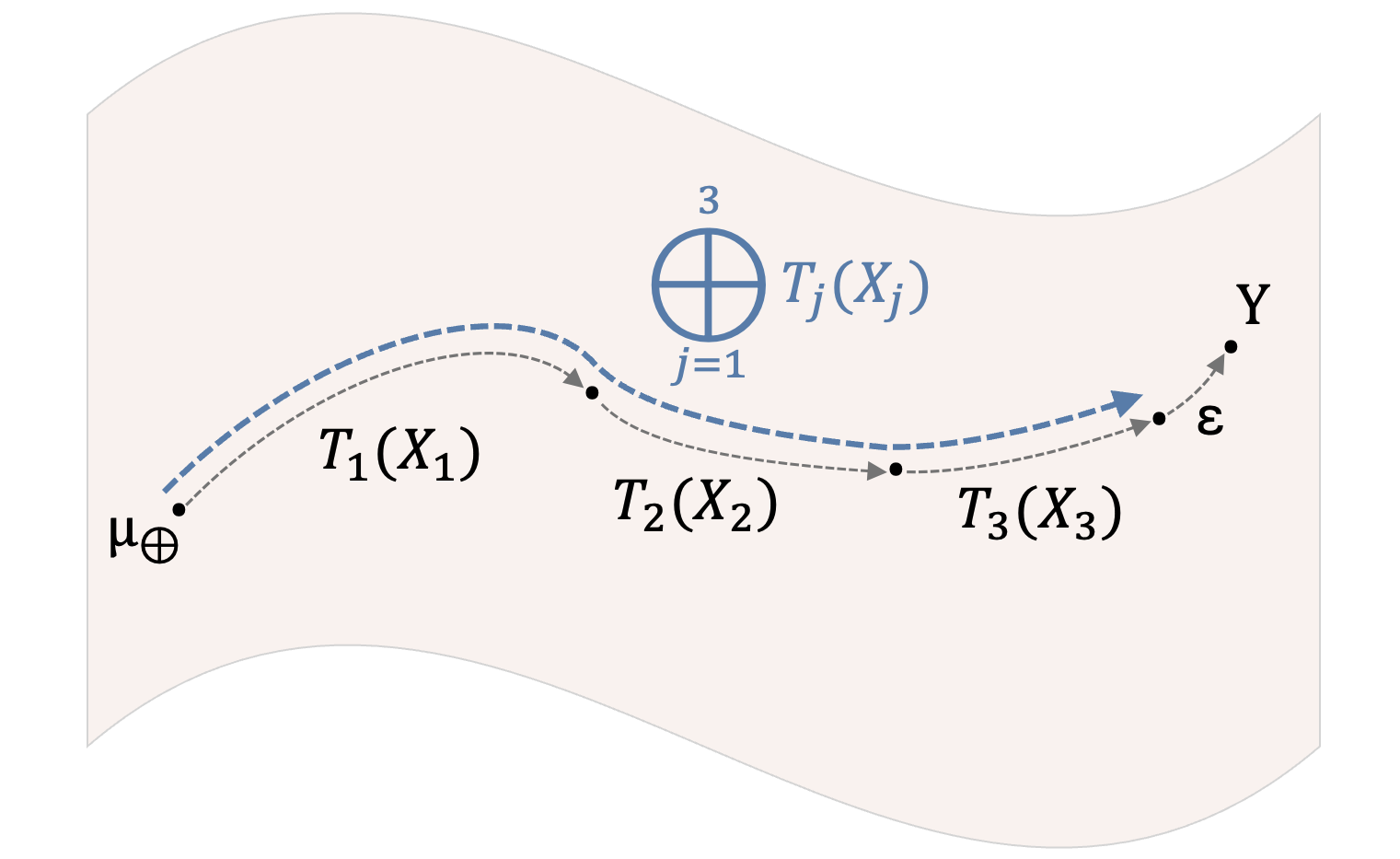}
     \includegraphics[width=0.45\linewidth]{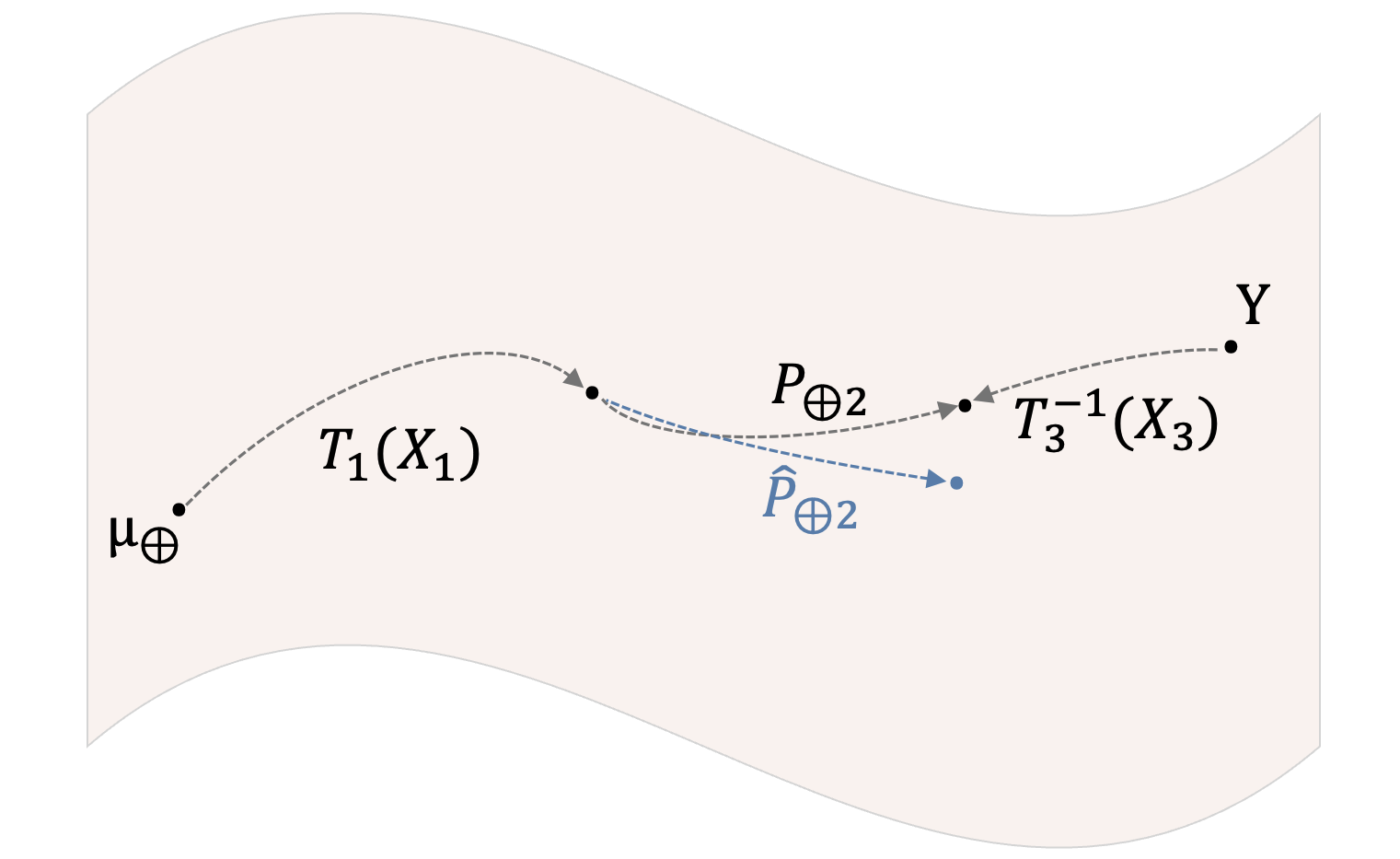}
    \caption{(Left panel) Additive optimal transport regression (ADOPT) framework with 3-dimensional predictors $X \in \real^{3}$ in \eqref{adopt_y}. (Right panel) Population partial transport residual $P_{\oplus j}$ in \eqref{eq:partial_tran_resi} and its estimator $\hat{P}_{\oplus j}$ in \eqref{eq:part_tran_resi_samp} with $j = 2.$ }
    \label{fig:framework}
\end{figure*}
We propose an additive optimal transport regression model that allows to regress $Y$ on the high dimensional predictors \textbf{X}. We first consider a scalar response $Y \in \mathbb{R}$ with predictors $\textbf{X} = \left(X_{1},\ldots, X_{p} \right).$ The classical additive regression model is the conditional mean of response $Y$ given $\bX = \bx,$ defined as
\begin{align}
    m(\bx) = \mathbb{E}[Y\mid \bX = \bx] = \beta_{0} + g_{1}(x_{1}) + \cdots + g_{p}(x_{p}),
    \label{addregscalar}
\end{align}
with the assumptions that $\mathbb{E}Y = \beta_{0},$ and $\mathbb{E}g_{j}(X_{j}) = 0,$ $j=1,\ldots, p,$ for identifiability. Then, the classical additive regression model \eqref{addregscalar} is equivalent to
\begin{align}
\begin{split}
     m(\bx) &= \left(g_{1}(x_{1}) - \mathbb{E}g_{1}(X_{1})\right) + \cdots \\
     &\quad \quad \quad \quad + \left(g_{p}(x_{p}) - \mathbb{E}g_{p}(X_{p})\right)  + \mathbb{E}Y.
\end{split}
    \label{addregscalarfinal}
\end{align}

In the case of scalar response $Y \in \mathbb{R}$ in (\ref{addregscalar}), the $j$th additive term $g_{j}(X_{j})$ is defined through conditional expectation of $Y$ given $X_{j} = x_j$ is as follows:
\begin{align*}
\begin{split}
     m_j(x_j) &= \mathbb{E}\left[Y \mid X_{j} = x_j\right] = \mathbb{E}\left[\mathbb{E}\left[Y \mid \bX \right] | X_{j} = x_j\right]  \\
     &= \mathbb{E}Y + g_{j}(x_{j}) + \mathbb{E}\left[\sum_{k\neq j} g_{k}(X_{k}) ~\bigg|~ X_{j} = x_j\right].
\end{split}
\end{align*}
Then, the additive term $g_{j}(X_{j})$ are given by 
\begin{align*}
    \begin{split}
         g_{j}(x_{j}) &= \mathbb{E}\left[P_{j}| X_{j} = x_{j} \right],
    \end{split}
\end{align*}
where for $j = 1,\ldots, p$  the $j$th partial residual is \begin{align}
    P_{j} = (Y - \mathbb{E}Y) - \sum\limits_{k\neq j} \left(g_{k}(X_{k}) - \mathbb{E}g_{k}(X_{k}) \right).
    \label{eq:partial_resi}
\end{align}  

A direct extension of model \eqref{addregscalarfinal} may not be feasible for responses situated in general metric spaces $\mathcal{M}$, since  basic vector space operations, such as addition, scalar multiplication and expectation may not exist. We generalize the expectation $\mathbb{E}$ to the \F mean $\mathbb{E}_{\oplus}$ in \eqref{eq:fm}. For convenience we reuse the notation for the $j$th additive term $g_j$,  
\begin{align*}
    g_{j}(x_{j}) \in \mathcal{M},\quad x_j \in \mathcal{X}_j, \quad j = 1, \ldots, p.
\end{align*}

The difference between the $j$th additive term and its \F mean 
is replaced by  the geodesic transport pushing $\mathbb{E}_{\oplus}g_{j}(X_{j})$ to $g_{j}(x_{j}),$ denoted as $T_{j}(x_{j}) := T_{\mathbb{E}_{\oplus}g_{j}(X_{j}),g_{j}(x_{j})} \in \mathcal{T}.$ Analogously, we introduce the $j$th inverse geodesic transport $T^{-1}_{j}(x_{j}) := T^{-1}_{\mathbb{E}_{\oplus}g_{j}(X_{j}),g_{j}(x_{j})} \in \mathcal{T}.$ 

We further note  that the actual responses in standard models  \eqref{addregscalar} are $Y=m(\bf{x}) + \varepsilon$, however in metric spaces additive errors $\varepsilon$ are not feasible. Contamination of responses by errors instead can be obtained through 
a random perturbation map $\M \rightarrow \M$ defined as $v'=\varepsilon(v)$,
such that  $\mathbb{E}_{\oplus}(v')=v$ and $\sigma^2=\mathbb{E}d^2(v,v')>0$
\citep{chen2022uniform}.
Assuming that the perturbation map is independent of predictors $\bX$, this 
leads to the proposed ADOPT model 
\begin{align}
    Y=\varepsilon(m_{\oplus}(\textbf{x})) = \left[\bigoplus_{j=1}^{p}T_{j}(x_{j}) \oplus \varepsilon \right](\mu_{\oplus}), \label{adopt_y}
\end{align} 
where $\mu_{\oplus} = \mathbb{E}_{\oplus}Y$ and $\bigoplus_{j=1}^{p}T_{j} = T_1 \oplus T_2 \oplus\cdots T_p.$

Generalizing partial residuals in the conditional expectation, the $j$th partial transport residual $P_{\oplus j} \in \mathcal{T},$ $j =1 ,\ldots, p,$ in the ADOPT model \eqref{adopt_y} is defined as
\begin{align}
    \begin{split}
        P_{\oplus j} &= \left[T^{-1}_{j-1}(X_{j-1}) \oplus \cdots \oplus T^{-1}_{1}(X_{1}) \right] \oplus T_{\mu_{\oplus},Y} \\
        &\quad\quad \oplus \left[T^{-1}_{p}(X_{p}) \cdots \oplus  T^{-1}_{j+1}(X_{j+1})\right].
    \end{split}
    \label{eq:partial_tran_resi}
\end{align}

The left panel of Figure \ref{fig:framework} shows the framework of the additive optimal transport regression and the right panel 
the partial transport residual $P_{\oplus 2}$ for a  3-dimensional predictor $\bX \in \real^{3}$.

\subsection{Estimation}

Given  i.i.d. pairs $\left(\textbf{X}_{i}, Y_{i} \right) \in \Pi_{j=1}^{p} \mathcal{X}_{j} \times \mathcal{M},$ where $\textbf{X}_{i} = \left(X_{i,1},\ldots, X_{i,p} \right),$  the sample ADOPT model is 
\begin{align}
    Y_{i} = \left[\bigoplus_{j=1}^{p}T_{j}(X_{i,j}) \oplus \varepsilon_{i} \right](\mu_{\oplus}), \quad i = 1,\ldots,n,
    \label{sampleadopt}
\end{align}
where we substitute the sample \F mean \eqref{eq:fm_samp} for  the \F mean $\mu_{\oplus}$, 
To fit model \eqref{sampleadopt}, we apply standard  backfitting \citep{breiman1985estimating, hastie1986generalized},  an  iterative algorithm with steps $t = 1,2, \ldots,$ alternating between 1) a backfitting step, where the partial residual $P_j$ in \eqref{eq:partial_resi} is smoothed with respect to $X_j,$ i.e., $g_j(x_j) \leftarrow \mathbb{E}(P_j\mid X = x_j),$ and 2) mean centering for identifiability, i.e., $g_j(x_j) \leftarrow g_j(x_j) - \mathbb{E}g_j(x_j),$ for all $j = 1, \ldots, p,$ until convergence. This alternating procedure is applied separately for each predictor, cycling through the predictor set. 
We initialize  with the \F mean estimate $\hat{\mu}_{\oplus}$ and the identity maps $\hat{T}^{(0)}_j(\cdot) = \mathrm{id}(\cdot) \in \mathcal{T},$ for $j = 1, 2, \ldots, p.$ 

At each iteration $t = 1, 2, \ldots$, we  construct partial transport residuals \begin{align}
\begin{split}
     &\hat{P}^{(t)}_{\oplus j} = \left[\left(\hat{T}_{j-1}^{(t)}\right)^{-1}(X_{j-1}) \oplus \cdots \oplus \left(\hat{T}_{1}^{(t)}\right)^{-1}(X_{1}) \right]  \oplus\\
        & T_{\hat{\mu}_{\oplus},Y} \oplus \left[\left(\hat{T}_{p}^{(t-1)}\right)^{-1}(X_{p}) \cdots \oplus  \left(\hat{T}_{j+1}^{(t-1)}\right)^{-1}(X_{j+1})\right],
\end{split}
   \label{eq:part_tran_resi_samp}
\end{align} with sample observations $\hat{P}^{(t)}_{\oplus i, j}$ obtained by replacing $(X_1, X_2, \ldots, X_p)$ with $(X_{i,1}, X_{i,2}, \ldots, X_{i,p}),$ $i = 1, 2, \ldots, n.$ The right panel of Figure \ref{fig:framework} illustrates the estimated partial transport residuals. 

The transport backfitting step utilizes local \F regression where the responses are the partial transport residuals, 
given each univariate predictor $X_j = x_j,$ \begin{align*}
\begin{split}
        g^{(t)}_{j}(x_j) &= \hat{\mathbb{E}}_{\oplus L}\left[\hat{P}^{(t)}_{\oplus j}(\hat{\mu}_{\oplus}) \mid  X_{j}=x_j\right]\\
        &= \argmin_{\up\in\mathcal{M}}\frac{1}{n}\sum_{i=1}^n\hat{w}_{i}(x_j, h)d^2\left(\hat{P}^{(t)}_{\oplus i, j}(\hat{\mu}_{\oplus}), \up\right).
\end{split}
\end{align*} The weights $\hat{w}_{i}$ are defined in \eqref{eq:loc_freg_samp} and the bandwidth $h$ is selected by using $5$-fold cross validation.

The mean centering step is generalized to a transport centering step, 
\begin{align*}
    \hat{T}^{(t)}_{j}(\cdot) = T_{\hat{\mathbb{E}}_{\oplus} g^{(t)}_{j}(X_j),\, g^{(t)}_{j}(\cdot)} \in \mathcal{T},
\end{align*} with sample \F mean 
\begin{align}
    \hat{\mathbb{E}}_{\oplus} g^{(t)}_{j}(X_j) = \argmin_{\up\in\mathcal{M}} \frac{1}{n} \sum_{i=1}^{n}d^2(g^{(t)}_{j}(X_{i,j}), \up).
    \label{eq:tran_center}
\end{align} 

At iteration $t,$ we apply the alternating steps cycling through predictors $X_j = x_j,$ $j = 1, \ldots, p$ and subsequently update the fitted responses  \begin{align*}
    \hat{Y}_i^{(t)} = \Big[\bigoplus\limits_{j=1}^{p}\hat{T}^{(t)}_{j}(X_{i,j})\Big](\hat{\mu}_{\oplus}), \quad i =1, 2, \ldots, n,
\end{align*} iterating until convergence when  \begin{align*}
    \frac{1}{n} \sum_{i=1}^{n}d^{2}(\hat{Y}_i^{(t)},\hat{Y}_i^{(t-1)}) < \epsilon,
\end{align*} for prespecified tolerance $\epsilon > 0.$ The workflow is summarized in Algorithm \ref{alg:transport_backfitting}.

Denote by $\hat{T}_j(\cdot)$  the estimates from Algorithm \ref{alg:transport_backfitting} when the other transports $T_{1}(\cdot),~ \ldots ,~T_{j-1}(\cdot),$ $T_{j+1}(\cdot), \ldots,~ T_{p}(\cdot)$ are assumed to be known. Then one can obtain the following oracle convergence under mild assumptions (see the Supplement Section A for the proof and detailed assumptions). 

\noindent \textbf{Theorem 1} \textit{Under Assumption 1 and the assumptions in  Supplement Section A, if the bandwidth $h \to 0$ and $nh \to \infty$ as $n \to \infty$, it holds that
\begin{align*}
    d\left(T_{j}(x_j)(\nu), \hat{T}_{j}(x_j)(\nu)\right) = o_{P}(1),
\end{align*} for any $x_j \in \mathcal{X}_j,$ and $\nu \in \M.$} 


\begin{algorithm}[h!]
    \caption{Transport Backfitting Algorithm}
    \label{alg:transport_backfitting}
    \begin{algorithmic}[1] 
    \REQUIRE $(X_i, Y_i) \in \real^{p} \times \M,$ $i = 1, 2, \ldots, n,$ and convergence tolerance $\epsilon > 0.$ \\
    \ENSURE $\hat{T}_{j}(\cdot) \in \mathcal{T},$ $j = 1, 2, \ldots, p.$
        \STATE Initialize  $\hat{T}^{(0)}_{j}(\cdot) = \mathrm{id},$ $j=1,\ldots,p$, and $\hat{Y}_i^{(0)} = \Big[\bigoplus\limits_{j=1}^{p}\hat{T}^{(0)}_{j}(X_{i,j})\Big](\hat{\mu}_{\oplus}),$  $i =1, 2, \ldots, n,$ with sample \F mean $\hat{\mu}_{\oplus}$.

        \FOR{$t = 1,2,\ldots$ until convergence}
            \FOR{$j = 1,\ldots,p$}
                \STATE Transport backfitting via local \F regression $\hat{\mathbb{E}}_{\oplus L}$ in \eqref{eq:loc_freg_samp},
                \[g^{(t)}_{j}(x_j) = \hat{\mathbb{E}}_{\oplus L}\left[\hat{P}^{(t)}_{\oplus j}(\hat{\mu}_{\oplus}) \mid  X_{j}=x_j\right],\] with partial transport residual observations $\hat{P}^{(t)}_{\oplus j}$ in \eqref{eq:part_tran_resi_samp}.
                \STATE Transport Centering,
                \[\hat{T}^{(t)}_{j}(\cdot) = T_{\hat{\mathbb{E}}_{\oplus} g^{(t)}_{j}(X_j),\, g^{(t)}_{j}(\cdot)} \in \mathcal{T},\] with \F mean $\hat{\mathbb{E}}_{\oplus} g^{(t)}_{j}(X_j)$ in \eqref{eq:tran_center}.
                \ENDFOR
            \STATE $\hat{Y}_i^{(t)} = \Big[\bigoplus\limits_{j=1}^{p}\hat{T}^{(t)}_{j}(X_{i,j})\Big](\hat{\mu}_{\oplus}),$ $i =1, 2, \ldots, n.$
            \STATE {Stopping rule:} Stop if \[\frac{1}{n} \sum_{i=1}^{n}d^{2}(\hat{Y}_i^{(t)},\hat{Y}_i^{(t-1)}) < \epsilon.\]
        \ENDFOR
    \end{algorithmic}
\end{algorithm}

\section{SIMULATIONS}

\subsection{Distribution-valued responses with Wasserstein metric}

We consider univariate distributional responses $Y$ situated in  the $2$-Wasserstein space $\mathcal{W}_2$ with Euclidean predictor vectors  $\bX \in \left[0,1\right]^{3},$ introduced in Example 1. In a slight abuse of notation, we use  $Y$ to denote the quantile function corresponding to a distribution. We generate distributional responses from the additive transport regression model in \eqref{sampleadopt}, \begin{align*}
    Y_i = \left[\bigoplus_{j=1}^{3} T_{j}(X_{i,j}) \oplus \varepsilon_i \right] \left(U \right),
\end{align*}
where $U = \text{Unif}(0,1),$  $T_{j}(x_{j}) = T_{\mathbb{E}_{\oplus}g_{j}(X_{j}), g_{j}(x_{j})},$ $x_j \in [0,1],$ $j = 1,2, 3,$ and $\varepsilon_i(f)(u) = f(u) + \frac{1}{2\pi}\xi_i\sin \left(2\pi f(u) \right),$ for all quantile functions $f \in \mathcal{W}_2,$ and $u \in \left[0,1\right],$ $\xi_i \sim \text{Unif}(-1,1)$. We consider the following two cases:
\begin{itemize}
    \item Case I (Beta distributions):  Here  $g_1(x_1) = \text{Beta}(1 + 2 x_1, 1),$ $g_2(x_2) = \text{Beta}(1, 2 + 3 x_2),$ and $g_3(x_3) = \text{Beta}(\frac{1}{2} + \frac{1}{2} x_3, \frac{1}{2} + \frac{1}{2} x_3).$
    \item Case II (Normal distributions): Here  $g_1(x_1) = N(x_1, 1),$ $g_2(x_2) = N(x_2^2, 1),$ and $g_3(x_3) = N(e^{-x_3}, 1).$
\end{itemize}
We generate 3-dimensional Euclidean predictors $\textbf{X}_{i} = (X_{i,1},X_{i,2},X_{i,3}) = \left(\Phi(V_{i,1}), \Phi(V_{i,2}), \Phi(V_{i,3})\right)^{T},$ $i = 1,\ldots, n,$ where $\Phi$ is the standard normal CDF and $\textbf{V}_{i} = (V_{i,1}, V_{i,2}, V_{i,3})^{T} \sim N_{3}(\textbf{0}, \Sigma),$ $\Sigma = \begin{pmatrix}
    1 & 0.5 & 0.3 \\
    0.5 & 1 & 0.5 \\
    0.3 & 0.5 & 1
\end{pmatrix}.$
For each case, we generate $n = 100,$ and $300$ pairs of $(\textbf{X}_{i},\textbf{Y}_{i}),$ $i = 1, 2, \ldots, n.$ We compare the finite-sample performance of the proposed ADOPT model with  global \F regression (GF) \citep{petersen2019frechet}, additive functional regression \citep{han2020additive} (ADR) and simply taking the constant \F mean (FM), in terms of the apporximated mean integrated squared error (MISE), 
\begin{align}
    \text{MISE} \approx B^{-1} \sum_{b=1}^{B} \int_{[0,1]^{3}} d^{2}(\hat{Y}_{\textbf{x}}^{(b)},Y_{\textbf{x}}) d\textbf{x},
    \label{eq:MISE}
\end{align}
where $d$ is 2-Wasserstein distance,  $\hat{Y}_{\textbf{x}}^{(b)}$ with $\bx = (x_1, x_2,x_3)$ are estimates from  the $b^{th}$ Monte Carlo (MC) samples $\left\{(\textbf{X}^{(b)}_{i},\textbf{Y}^{(b)}_{i}) : 1 \leq i \leq n \right\},$ and $Y_{\textbf{x}}$ are the true distributional data without noise,  $Y_{\textbf{x}} = \left[\bigoplus_{j=1}^{p} T_{j}(x_{j}) \right] \left(U\right).$ The results are in 
Table \ref{tbl1}   
and demonstrate that 
ADOPT has the smallest MISE among the comparison methods and  as $n$ increases the MISE of ADOPT decreases,  while this is not the case for the competing  methods. 

\begin{table}[h!]
\caption{Mean integrated squared error (MISE) results  ($\times 10^{-3}$) with standard errors in parentheses  under distributional responses in the Wasserstein space for two cases  with different sample sizes $n = 100,$ and $300$. We compare four methods: additive optimal transport regression (ADOPT), global \F regression \citep{petersen2019frechet} (GF), additive functional regression \citep{han2020additive} (ADR), and \F mean \citep{frec:48} (FM) without predictor effect.}
\label{tbl1}
\centering
\begin{tabular}{cc *{4}{S[table-format=2.3]}}
\toprule
& & \multicolumn{4}{c}{\textbf{Method}} \\
\cmidrule(lr){3-6}
{Case} & ${n}$ & {ADOPT} & {GF} & {ADR} & {FM} \\
\midrule
\multirow{4}{*}{I} & \multirow{2}{*}{100} & \textbf{~0.317} & 0.591 & 0.761 & 11.019 \\
& & \se{0.010} & \se{0.008} & \se{0.008} & \se{0.068} \\
\addlinespace
& \multirow{2}{*}{300} & \textbf{~0.181} & 0.499 &  0.586 & 11.119 \\
& & \se{0.004} & \se{0.003} & \se{0.004} & \se{0.040} \\
\midrule
\multirow{4}{*}{II} & \multirow{2}{*}{100} & \textbf{~1.685} & 6.636 & 80.411 & 207.205 \\
& & \se{0.057} & \se{0.057} & \se{0.702} & \se{1.681} \\
\addlinespace
& \multirow{2}{*}{300} & \textbf{~1.121} & 6.701 & 81.951 & 211.297 \\
& & \se{0.041} & \se{0.037} & \se{0.455} & \se{1.080} \\
\bottomrule
\end{tabular}
\end{table}

\subsection{SPD matrices-valued response with log-Cholesky metric}

We consider $m \times m$ SPD matrix valued responses $Y \in \mathcal{S}_{m}^{+}$ equipped with the log-Cholesky metric \citep{lin2019riemannian} introduced in Example 2, and Euclidean predictors $\bX \in [0,1]^3$ and generate SPD responses from the additive transport regression model in \eqref{sampleadopt},
\begin{align*}
    Y_i = \left[ \bigoplus_{j=1}^{3} T_{j}(X_{i, j}) \oplus \varepsilon_i \right] (\mathbf{I}_{m}) \in \mathcal{S}^{+}_{m},
\end{align*}
where $\mathbf{I}_{m}$ is  $m \times m$ identity matrix, $T_{j}(x_{j}) = T_{\mathbb{E}_{\oplus}g_{j}(X_{j}),g_{j}(x_{j})}, $ $j = 1,\ldots, 3,$ and for $S = LL^{\mathrm{T}} \in \mathcal{S}^+_{m}$ with Cholesky decomposition, the perturbation error map $\varepsilon_i(S) = \varepsilon_i(LL^{\mathrm{T}}) = e_i(L) e_i(L)^\mathrm{T},$ where $e_i(L)_{k,l} = L_{k,l} + \xi_{k,l},$ and $\xi_{k,l} \sim N(0, \sigma^2),$ $k > l,$ $e_i(L)_{k,l} = L_{k,l},$ when $k = l,$ and $e_i(L)_{k,l} =0$ if $k < l.$ 

We generate the transport map $T_j(x_j)$ based on \begin{align*}
    g_j(x_j) = L_j(x_j) L_j(x_j)^{\mathrm{T}},\quad x_j \in [0,1], \quad j = 1,2,3,
\end{align*} with random Cholesky factors $L_j(x_j)$ which are  lower triangular matrices as follows: 
\begin{itemize}
    \item Case I (Linear in $\bx$): \begin{align*}
        \left(L_{1}(x_{1})\right)_{k,l} &= \exp^{-|k-l|/2}\left(x_1 + 1\right)/2, \\
        \left(L_{2}(x_{2})\right)_{k,l} &= \exp^{-|k-l|/3}\left(x_2 + 1\right)/4, \\
        \left(L_{3}(x_{3})\right)_{k,l} &= \exp^{-|k-l|/4}\left(x_3 + 1\right)/8.
    \end{align*}
    \item Case II (Non-linear in $\bx$): \begin{align*}
        \left(L_{1}(x_{1})\right)_{k,l} &= \exp^{-|k-l|/2}\sin\left((x_1+1)\pi/4 \right), \\
        \left(L_{2}(x_{2})\right)_{k,l} &= \exp^{-|k-l|/4}\left(x^2_2 + 1\right)/2, \\
        \left(L_{3}(x_{3})\right)_{k,l} &= \exp^{-|k-l|/4}\exp^{-x_3},
    \end{align*}
\end{itemize} for $k \ge l,$ and all other elements are $0$ if $k < l.$

We then generate random covariates $\textbf{X}_{i} = (X_{i,1},X_{i,2},X_{i,3}),$ $i = 1,\ldots, n,$ following the same procedure as in the distribution-valued simulations of Section 5.1. The responses are generated under each scenario with perturbations $\varepsilon_i$ having standard deviation $\sigma = 0.01.$ We sample $n=100$ and $300$ i.i.d. pairs of $(\textbf{X}_{i}, Y_{i}) \in \mathbb{R}^{3} \times \mathcal{S}^{+}_{m},$ with matrix sizes $m =  10$ and $m=20$ ands compare the finite sample performance of the proposed ADOPT with the global \F regression (GF) and the constant \F mean (FM) in terms of mean integrated squared error (MISE), analogously defined as in \eqref{eq:MISE}.
Table \ref{tbl2} displays the MISEs for $B= 200$ with standard error in parentheses. ADOPT has the lowest MISE and its MISE decreases as the sample size $n$ increases, while this is not the case for the  competing  methods. 
As the size of SPD matrices increases, the MISE increases, but ADOPT always has substantially better performance. 

\begin{table}[h!]
\caption{Mean integrated squared error (MISE) results  ($\times 10^{-2}$) with standard errors in parentheses for $m \times m$ SPD matrices as responses in $\mathcal{S}_{m}^{+}$ with the log-Cholesky metric for two simulation cases with different sample sizes $n = 100,$ and $300$ and matrix dimensions $m = 10,$ and $20$, comparing three methods: additive optimal transport regression (ADOPT), global \F regression  (GF) and \F mean (FM).}
\label{tbl2}
\centering
\begin{tabular}{ccc *{3}{S[table-format=2.3]}}
\toprule
& & & \multicolumn{3}{c}{\textbf{Method}} \\
\cmidrule(lr){4-6}
{Case} & ${m}$ & $n$ & {ADOPT} & {GF} & {FM} \\
\midrule
\multirow{8}{*}{I} & \multirow{4}{*}{10} & \multirow{2}{*}{100}  & \textbf{~2.769} & 9.495 & 135.344 \\
& &  & \se{0.025} & \se{0.060} & \se{0.647} \\
\addlinespace
&  & \multirow{2}{*}{300} & \textbf{~1.655}& 9.714 & 136.122 \\
& &  & \se{0.012} & \se{0.037} & \se{0.366} \\
\addlinespace
 & \multirow{4}{*}{20} & \multirow{2}{*}{100}  & \textbf{~5.712} & 14.110 & 192.688 \\
& &  & \se{0.070} & \se{0.102} & \se{0.92} \\
\addlinespace
&  & \multirow{2}{*}{300} & \textbf{~3.788} & 14.254 & 193.796 \\
& &  & \se{0.048} & \se{0.063} & \se{0.521} \\
\midrule
\multirow{8}{*}{II}  & \multirow{4}{*}{10} & \multirow{2}{*}{100}  & \textbf{~4.146} & 18.507 & 96.562 \\
& &  & \se{0.088} & \se{0.090} & \se{0.490} \\
\addlinespace
&  & \multirow{2}{*}{300} & \textbf{~2.702} & 18.808 & 97.047 \\
& &  & \se{0.036} & \se{0.052} & \se{0.303} \\
\addlinespace
 & \multirow{4}{*}{20} & \multirow{2}{*}{100}  & \textbf{~8.921} & 28.237 & 143.382 \\
& &  & \se{0.167} & \se{0.14} & \se{0.706} \\
\addlinespace
&  & \multirow{2}{*}{300} & \textbf{~6.435} & 28.512 & 144.088 \\
& &  & \se{0.100} & \se{0.078} & \se{0.437} \\
\bottomrule
\end{tabular}
\end{table}

\section{BRAIN CONNECTIVITY ANALYSIS}

We use resting state functional Magnetic Resonance Imaging (rs-fMRI) data obtained from the Alzheimer's Disease Neuroimaging Initiative (ADNI) database (\url{adni.loni.usc.edu}). Brain signal analysis at the subject level relies  on time series of Blood Oxygen Level Dependent (BOLD) signals, obtained for  a set of regions of interest (ROIs). The coherence between pairwise ROIs is customarily quantified  by Pearson correlation coefficients of the rs-fMRI time series, leading to a $l \times l$ correlation matrix for $l$ ROIs \citep{badhwar2017resting, perovnik2023functional}. {
The BOLD signals for each subject were collected over a time interval from 0 to 270 seconds with $K = 136$ equispaced measurements taken every 2 seconds. We have $136 \times 11$ signal matrix $S$, and $s_{kl}$ is the $(k,l)$th element of the signal matrix $S$ and an $11 \times 11$ connectivity matrix response $Y$ was obtained using Pearson correlation,  \begin{align*}
    (Y)_{l_1,l_2} = \frac{\sum_{k=1}^{K} (s_{kl_1} - \bar{s}_{l_1})(s_{kl_2} - \bar{s}_{l_2})  }{\sqrt{\sum_{k=1}^{K} (s_{kl_1} - \bar{s}_{l_1})^2} \sqrt{\sum_{k=1}^{K} (s_{kl_2} - \bar{s}_{l_2})^2}},
\end{align*} with $\bar{s}_{l} = \frac{1}{K}\sum_{k=1}^{K}s_{kl},$ $l_1, l_2 = 1,2, \ldots, 11.$ For Alzheimer’s disease (AD) studies, the CSF phosphorylated tau (p-Tau) concentration is another established biomarker, where elevated pTau levels are strongly associated with AD pathology \citep{karikari2022blood}, with higher p-Tau typically indicating greater AD risk. Also, cerebrospinal fluid (CSF) amyloid-beta (A$\beta$) concentration is a widely used marker of amyloid pathology \citep{murphy2010alzheimer, hampel2021amyloid}.} Further details about ADNI data preprocessing, and the list of $l = 11$ ROIs used in the analysis \citep{andrews2010functional} are described in Supplement Section B.1.

We analyze the $11 \times 11$ correlation matrices equipped with the log-Cholesky metric as responses using the ADOPT model in \eqref{sampleadopt}, based on $n = 929$ ADNI participants classified into six diagnostic stages: 248 cognitively normal (CN), 110 subjective memory complaint (SMC), 316 early mild cognitive impairment (EMCI), 12 mild cognitive impairment (MCI), and 171 late mild cognitive impairment (LMCI), and 72 Alzheimer's disease (AD). 


We consider $p=3$ predictors: $X_1$ is A$\beta$ concentration ranging from 203 to 1700, with lower amyloid-beta signaling amyloid-burden; $X_2$ is the diagnostic stage coded ordinally from $0$ to $5$ in the order CN, SMC, EMCI, MCI, LMCI, AD;  $X_3$ is the p-Tau concentration ranging  from $8.00$ to $92.08,$ with higher p-Tau typically indicating worse prognosis.  

To compare brain coherence patterns across distinct risk profiles, we construct three covariate settings representing low, intermediate, and high risk groups. The low risk group corresponds to the upper 10\% of A$\beta$ values (1700), the lower 10\% disease stage (CN = 0), and p-Tau values (12.8). The intermediate risk group is defined by median values of A$\beta$ (995.8), disease stage (EMCI = 2), and p-Tau (21.94). The high risk group corresponds to the lower 10\% of A$\beta$ values (535), the upper 90\% disease stage (LMCI = 4), and p-Tau values (42.01). 

\begin{figure*}[!h]
    \centering
    \includegraphics[width=0.75\linewidth]{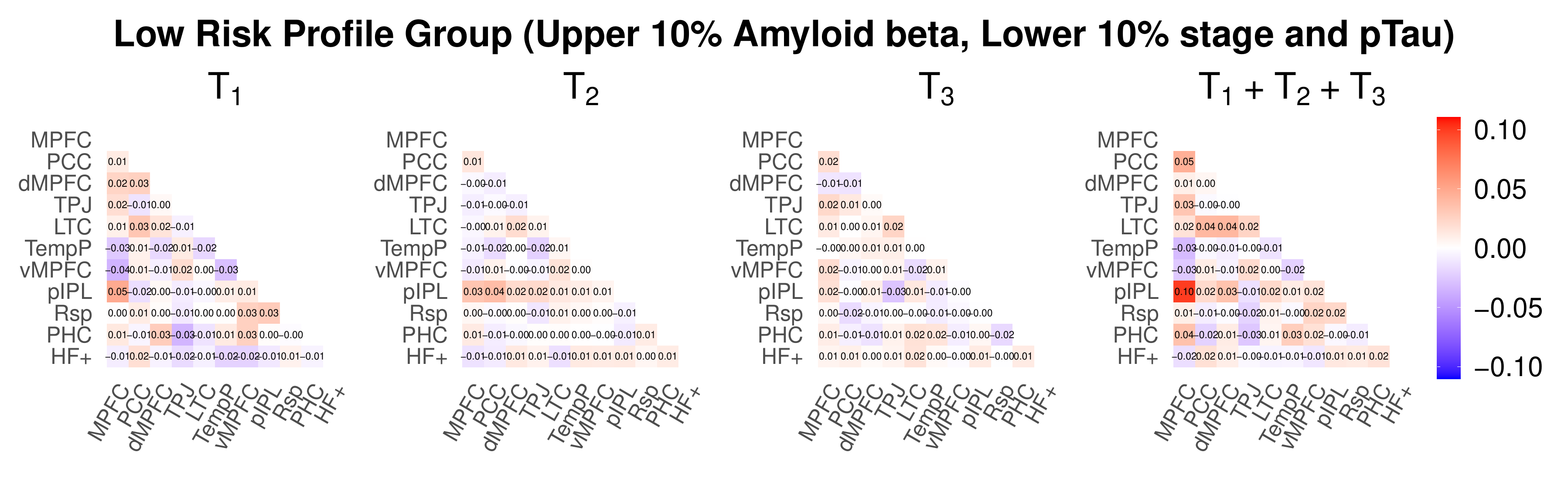}
    \includegraphics[width=0.75\linewidth]{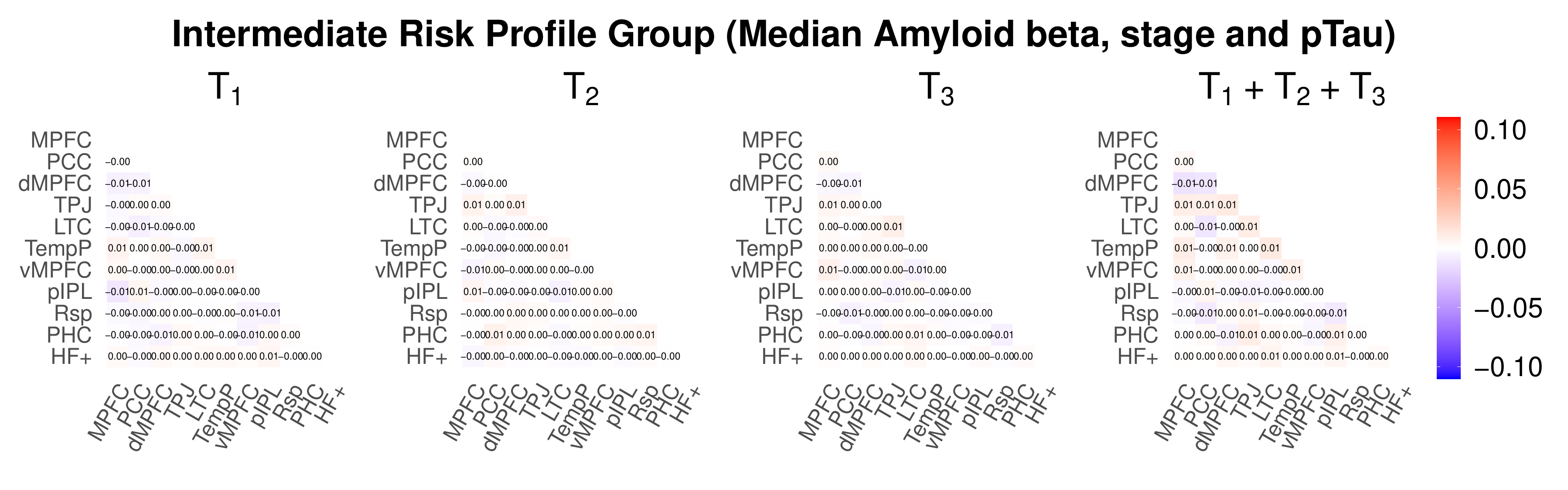}
    \includegraphics[width=0.75\linewidth]{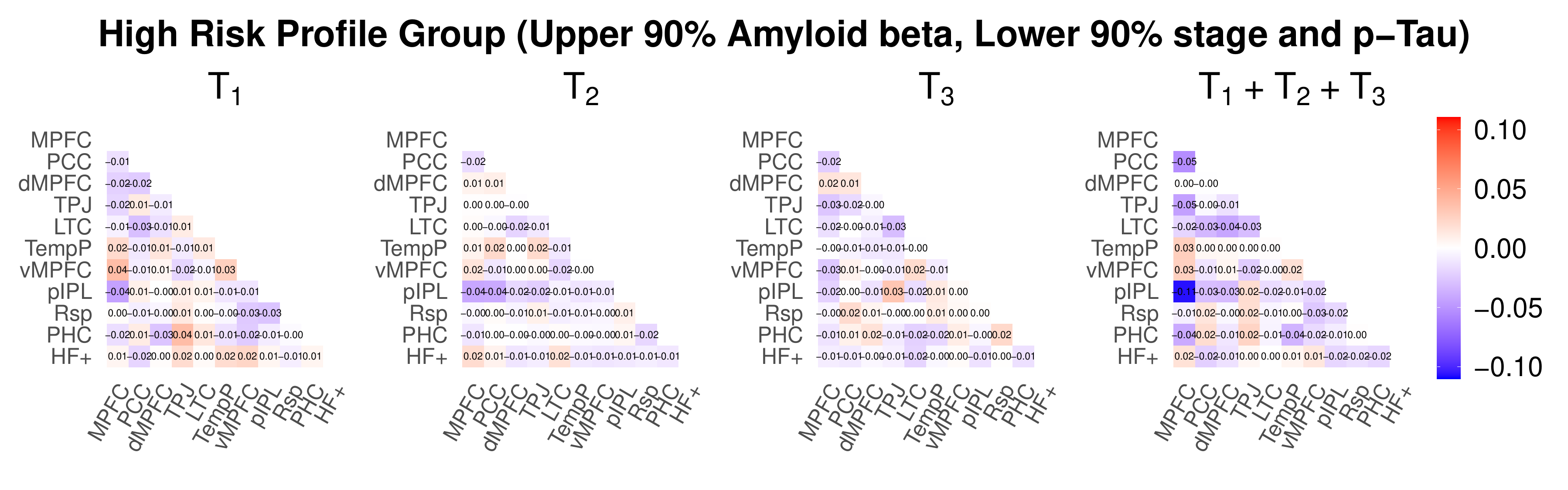}
    \caption{Estimated transport maps ${T}_1(x_1)$ (first column), ${T}_2(x_2)$ (second column), ${T}_3(x_3)$ (third column), and their combined map $[{T}_1(x_1) \oplus {T}_2(x_2) \oplus {T}_3(x_3)]$ (fourth column) represented as the difference of the lower triangular Cholesky factors, with regions of interest (ROIs) labeling rows and columns of each Cholesky factor. Rows correspond to low, intermediate, and high risk groups, with covariates $\bx=(x_1,x_2,x_3)$ defined as follows: the low risk group corresponds to the upper 10\% of A$\beta$ values (1700), the lower 10\% disease stage (CN = 0), and p-Tau values (12.8); the intermediate group is defined by the median A$\beta$ value (995.8), median disease stage (EMCI = 2), and median p-Tau value (21.94); and the high risk group corresponds to the upper 90\% of A$\beta$ values (535), the lower 90\% disease stage (LMCI = 4), and p-Tau values (42.01).}
    \label{fig:adni_transport}
\end{figure*}

Figure~\ref{fig:adni_transport} displays each transport map $T_j(x_j) = T_{\mathbb{E}_{\oplus}g(X_j), g(x_j)}$ as the difference of the Cholesky factors of the two correlation matrices, i.e., $L_{g(x_j)} - L_{\mathbb{E}_{\oplus}g(X_j)},$ where $L_S$ is a lower triangular Cholesky factor of SPD matrices $S \in \mathcal{S}_{l}^{+}$, with ROIs labeling rows and columns of each Cholesky factor. The estimated transport maps ${T}_1(x_1)$ (first column), ${T}_2(x_2)$ (second column), ${T}_3(x_3)$ (third column), and their combined effect $[{T}_1(x_1) \oplus {T}_2(x_2) \oplus {T}_3(x_3)]$ (fourth column) are obtained from ADOPT. The rows correspond to the low, intermediate, and high risk groups, with covariates $\bx=(x_1,x_2,x_3)$ defined in the previous paragraph. For the low risk group (first row), the transport maps generally indicate increasing correlations between ROIs, with the largest increase of 0.10 observed between pIPL and MPFC. For the intermediate group (second row), the transport maps are close to zero, implying that the \F mean $\hat{\mu}_{\oplus}$ is similar to the estimated response $[{T}_1 \oplus {T}_2 \oplus {T}_3 ](\hat{\mu}_{\oplus})$. In contrast, the high risk group (third row) shows decreasing correlations, with the strongest decline of 0.11 between pIPL and MPFC.

{The additive structure of the model leads to enhanced interpretability, providing  predictor-specific transport maps $T_1, T_2,$ and $T_3$  corresponding to the individual effects of $x_1$ (A$\beta$), $x_2$ (disease stage), and $x_3$ (p-Tau), respectively. For the disease-stage transport map $T_2,$ the strongest effect appears in the connection between the posterior cingulate cortex (PCC) and the posterior inferior parietal lobule (pIPL): approximately $+0.04$ in the low disease stage group versus $-0.04$ in the high disease stage group, consistent with well-known findings of reduced PCC–medial temporal lobe region (MTL, including pIPL) connectivity in \cite{andrews2010functional}. For the p-tau transport map $T_3,$ we find an overall decrease in connectivity in the high-risk group, consistent with literature showing that elevated p-tau is associated with broad reductions in functional connectivity \citep{luan2024tau}.}

We evaluate out-of-sample prediction performance using 5-fold cross-validation. Specifically, we partition the index set ${1,\ldots,n}$ into disjoint folds $\mathcal{I}_q$ $(q=1,\ldots,5)$ with $n_q = |\mathcal{I}_q|$. For each fold, we define the Mean Prediction Error Reduction (MPER) as
\begin{align*}
    \text{MPER} &=  \frac{1}{5}\sum_{q=1}^{5} \text{MPER}_{q}, \\
    \text{MPER}_{q} &= \frac{1}{n_{q}} \sum_{i \in \mathcal{I}_q} \left\{ d(\hat{\mu}_{\oplus}^{(-q)}, Y_i) - d(\hat{Y}^{(-q)}_i, Y_i) \right\},
\end{align*}  where
$\hat{Y}_i^{(-q)}$ is the predicted response from the regression model trained on the remaining folds, $\hat{\mu}_{\oplus}^{(-q)}$ is the \F mean of the training responses (baseline without predictors) and $d$ is the log-Cholesky metric. MPER measures the reduction in prediction error relative to the baseline; larger values indicate better predictive performance. We compare our method with global \F regression (GF) \citep{petersen2019frechet}. {Table \ref{tbl3} reports 
the average of MPER $(\times 10^{-4})$  for  $B=200$ Monte Carlo simulations  for ADOPT and global \F regression, demonstrating  that ADOPT improves out-of-sample performance substantially in terms of MPER.
Additional simulation results are provided in Supplement Section B.2.}

\begin{table}[h!]
\caption{Mean Prediction Error Reduction (MPER, $\times 10^{-4}$) with standard errors in parentheses, based on ADNI brain correlation regression analysis with log-Cholesky metric. Higher MPER values indicate better out-of-sample prediction. We compare two methods: additive optimal transport regression (ADOPT), and global \F regression \citep{petersen2019frechet} (GF)}
\label{tbl3}
\centering
\begin{tabular}{ccc}
\toprule
&\textbf{ADOPT} & \textbf{GF} \\
\midrule
MPER & \textbf{2.625 (0.0554)}  & 1.713 (0.0518) \\
\bottomrule
\end{tabular}
\end{table}

\section{DISCUSSION}

We develop ADOPT, an additive optimal transport regression approach for random object responses with multivariate continuous Euclidean predictors, thereby providing a flexible regression model for metric space-valued responses that addresses  the curse of dimensionality and enhancing interpretability in real data analysis. The method is built on an extension of optimal transport ideas to general metric spaces and utilizes a  transport map for  each predictor. 

We illustrate the proposed ADOPT with brain connectivity for resting state fMRI brain imaging data. Numerical experiments with distributional responses in Wasserstein space and SPD matrices with the log-Cholesky metric also demonstrate strong performance. 

\subsubsection*{Acknowledgements}
We acknowledge the helpful and valuable comments from anonymous reviewers, which led to numerous improvements. This research was supported in part by NSF grant DMS-2310450. This article is motivated by exploring the association between neuroimaging measures and cognitive function in patients with Alzheimer’s disease (AD) through the analysis of the Alzheimer’s Disease Neuroimaging Initiative (ADNI) study (http://www.adni-info.org/). The ADNI is a large-scale multi-site neuroimaging study that has collected clinical, imaging, genetic and cognitive data at multiple time points from cognitive normal (CN) subjects, subjects with mild cognitive impairment (MCI), and AD patients.


\bibliography{mybibfile.bib}
\section*{Checklist}



\begin{enumerate}

  \item For all models and algorithms presented, check if you include:
  \begin{enumerate}
    \item A clear description of the mathematical setting, assumptions, algorithm, and/or model. [Yes] Assumptions are introduced in Section 2 and Supplement Section A, the proposed model and algorithm for estimation are provided in Section 4.
    \item An analysis of the properties and complexity (time, space, sample size) of any algorithm. [Yes] We provide consistency results in Section 4, and complexity analysis in Supplement Section C.
    \item (Optional) Anonymized source code, with specification of all dependencies, including external libraries. [Partial Yes] The resting-state fMRI data we used are fully accessible to qualified researchers through the ADNI website (https://adni.loni.usc.edu) upon approval of a standard data-use application (https://adni.loni.usc.edu/data-samples/adni-data). We cannot redistribute the raw data ourselves due to ADNI’s data-use agreement, not meaning that the data are unavailable to other researchers.
  \end{enumerate}

  \item For any theoretical claim, check if you include:
  \begin{enumerate}
    \item Statements of the full set of assumptions of all theoretical results. [Yes] All theoretical assumptions are provided in Supplement Section A.
    \item Complete proofs of all theoretical results. [Yes] All theoretical proofs are provided in Supplement Section A.
    \item Clear explanations of any assumptions. [Yes] See Supplement Section A.    
  \end{enumerate}

  \item For all figures and tables that present empirical results, check if you include:
  \begin{enumerate}
    \item The code, data, and instructions needed to reproduce the main experimental results (either in the supplemental material or as a URL). [Partial Yes] The resting-state fMRI data we used are fully accessible to qualified researchers through the ADNI website (https://adni.loni.usc.edu) upon approval of a standard data-use application (https://adni.loni.usc.edu/data-samples/adni-data).  We cannot redistribute the raw data ourselves due to ADNI’s data-use agreement, not meaning that the data are unavailable to other researchers.
    \item All the training details (e.g., data splits, hyperparameters, how they were chosen). [Yes]
    \item A clear definition of the specific measure or statistics and error bars (e.g., with respect to the random seed after running experiments multiple times). [Yes]  
    \item A description of the computing infrastructure used. (e.g., type of GPUs, internal cluster, or cloud provider). [Yes] We use 60 CPUs for the simulation analysis.
  \end{enumerate}

  \item If you are using existing assets (e.g., code, data, models) or curating/releasing new assets, check if you include:
  \begin{enumerate}
    \item Citations of the creator If your work uses existing assets. [Not Applicable]
    \item The license information of the assets, if applicable. [Not Applicable]     
    \item New assets either in the supplemental material or as a URL, if applicable. [Not Applicable]
    \item Information about consent from data providers/curators. [Not Applicable] 
    \item Discussion of sensible content if applicable, e.g., personally identifiable information or offensive content. [Not Applicable]
  \end{enumerate}

  \item If you used crowdsourcing or conducted research with human subjects, check if you include:
  \begin{enumerate}
    \item The full text of instructions given to participants and screenshots. [Not Applicable]
    \item Descriptions of potential participant risks, with links to Institutional Review Board (IRB) approvals if applicable. [Not Applicable]
    \item The estimated hourly wage paid to participants and the total amount spent on participant compensation. [Not Applicable]
  \end{enumerate}

\end{enumerate}

\clearpage
\appendix
\thispagestyle{empty}

\onecolumn
\aistatstitle{Supplementary Materials to \\
``ADOPT: Additive Optimal Transport Regression"}

\setcounter{equation}{0}
\setcounter{table}{0}

\renewcommand{\theequation}{S.\arabic{equation}}
\renewcommand{\thetable}{S.\arabic{table}}

\section{Technical Details}

\subsection{Assumptions}

In this section, we state the assumptions required for Theorem 1 and introduce the necessary notation. For all $\nu \in \M$ and $T_j \in \mathcal{T},$ $j = 1, \ldots, p,$ define $Z_j(x_j) = \left[ \varepsilon \circ T_j (x_j) \right] (\nu) \in \M,$ $x_j \in \mathcal{X}_j,$ where $\varepsilon \in \mathcal{T}$ is a random perturbation map \citep{chen2022uniform}. We define $T_j(x_j) = T_{\mathbb{E}_{\oplus}g_j(X_j), g_j(x_j)} \in \mathcal{T},$ and  \begin{align*}
    M_{\oplus j}(\omega, x_j; \nu) &= \mathbb{E} \left[ d^2\left(Z_j(X_j)(\nu), \omega\right) \mid X_j = x_j \right], \\
    M^L_{\oplus j}(\omega, x_j; \nu) &= \mathbb{E}\left[w(x_j, h) d^{2}\left(Z_j(X_j)(\nu), \omega\right) \right], \\
    \hat{M}^L_{\oplus j}(\omega, x_j; \nu) &= \frac{1}{n} \sum_{i=1}^n \hat{w}_i(x_j, h) d^{2}\left(Z_{i,j}(X_{i,j})(\nu), \omega\right),
\end{align*} where $Z_{i,j}(X_{i,j})(\nu) = \left[\varepsilon_i \circ T_j(X_{i,j}) \right] (\nu) \in \M.$

By definition we have \begin{align*}
    T_j(x_j)(\nu) &= \argmin_{\omega \in \M} M_{\oplus j}(\omega, x_j; \nu), \\
    T^L_j(x_j)(\nu) &= \argmin_{\omega \in \M} M^L_{\oplus j}(\omega, x_j; \nu), \\
    \hat{T}^L_j(x_j)(\nu) &= \argmin_{\omega \in \M} \hat{M}^L_{\oplus j}(\omega, x_j; \nu).
\end{align*}

We need the following assumptions:
\begin{enumerate}
    \item[\textit{(A1)}] The kernel function $K$ is a symmetric probability density function centered at zero. Furthermore, define $K_{kl} = \int_{\real} K^{k}(u) u^{j} du,$ where $\lvert  K_{14} \rvert$ and $\lvert K_{26} \rvert$ are both finite.

    \item[\textit{(A2)}] For all $j = 1, \ldots, p,$ the marginal density $f_j$ of $X_j$ as well as the conditional densities $f_{\mid z_j}(x_j)$ of $X_j \mid Z_j(x_j)(\nu) = z_j$ and $f_{\mid x_j}(z_j)$ of $Z_j \mid X_j = x_j$ exist. The conditional densities $f_{\mid z_j}(x_j)$ are twice continuously differentiable for all $z_j \in \M$ and $\sup_{x_j,z_j}\lvert f{''}_{\mid z_j}(x_j) \rvert < \infty.$ For any open set $U \subset \M,$ $\int_{U} d F_{\mid x_j}(z_j)$ is continuous in $x_j \in \mathcal{X}_j,$ where $F_{\mid x_j}(z_j)$ is the cumulative distribution function corresponding to $f_{\mid x_j}(z_j).$

    \item[\textit{(A3)}] For all $\nu \in \M,$ and $x_j \in \mathcal{X}_j,$ $j = 1, \ldots, p,$ objects $T_j(x_j)(\nu),$ $T^L_j(x_j)(\nu),$ and $\hat{T}^L_j(x_j)(\nu)$ exist and are unique, the latter almost surely. For all $\delta > 0,$ \begin{align*}
        \inf_{d(\omega, T_j(x_j)(\nu) ) < \delta} \left[M_{\oplus j}(\omega, x_j; \nu) - M_{\oplus j}(T_j(x_j)(\nu), x_j; \nu) \right] > 0, \\
        \liminf_{h \to 0}\inf_{d(\omega, T^L_j(x_j)(\nu) ) < \delta} \left[M^L_{\oplus j}(\omega, x_j; \nu) - M^L_{\oplus j}(T^L_j(x_j)(\nu), x_j; \nu) \right] > 0.
    \end{align*}

\end{enumerate}

Condition \textit{(A1)} is standard kernel requirements typically assumed in local regression estimation. Condition \textit{(A2)} is distributional assumptions on the predictors and responses for the convergence of \F regression estimators \citep{petersen2019frechet}. Condition \textit{(A3)} is a regularity condition commonly used to establish the consistency of M-estimators \citep{van2023weak}.

We need the following assumption on the geodesic optimal transport:
\begin{enumerate}
    \item[\textit{(A4)}] (Small Error Assumption) For perturbation maps $\left\{\varepsilon_{i,n} \right\}_{i=1}^{n}$ with sample size $n,$ there exists a sequence $\delta_n >0$ with $\delta_n \to 0$ as $n \to \infty,$ such that \begin{align*}
        \mathbb{E} \left[d(\varepsilon_{1,n}(\nu), \nu) \right] \le \delta_n,
    \end{align*} for all $\nu \in \M.$
    \item[\textit{(A5)}] There exists a constant $C >0$ such that \begin{align*}
        d(\Gamma(\up_1, \up_2, \up_3), \Gamma(\up'_1, \up'_2, \up'_3)) \le C\left\{d(\up_1, \up'_1) + d(\up_2, \up'_2) + d(\up_3, \up'_3)\right\}
    \end{align*}
    \item[\textit{(A6)}] The transport maps $T_j(\cdot) \in \mathcal{T}$ are perturbation maps, i.e., for all $\nu \in \M,$ \begin{align*}
        \nu = \argmin_{\omega \in \M} \mathbb{E}\left\{d^2(T_j(X_j)(\nu), \omega) \right\},
    \end{align*} and \begin{align*}
        \inf_{d(\omega, \nu) > \delta} \left[ \mathbb{E}d^2(T_j(X_j)(\nu), w) - \mathbb{E}d^2(T_j(X_j)(\nu),\nu) \right] > 0,
    \end{align*} for any $\delta > 0.$
\end{enumerate}
{The small error assumption (A4) appears at various instances in the classical nonparametric regression literature, including generalized regression models for longitudinal data \citep{hall2008modelling}, errors-in-variables regression \citep{lee2023depth}, and Tikhonov/RKHS-regularized regression \citep{lang2023small}. (A4) is only required for Lemma 3, where we need the quantity $d(T \oplus \varepsilon,\varepsilon \oplus T)$ to be small, because the composition of the transport map with the error-perturbation map is not commutative and our transport backfitting algorithm iteratively applies partial transport maps. Controlling this non-commutativity error is therefore essential for establishing the theoretical guarantees. We note that (A4)  is not needed in metric spaces where the $\oplus$ operations are commutative (e.g., SPD matrices under the log–Cholesky metric). }


Condition \textit{(A5)} holds for random objects in complete, non-positively curved metric spaces (Hadamard spaces) \citep{zhu2025geodesic}. Examples include the space of univariate distributions with the Wasserstein metric and the space of symmetric positive definite (SPD) matrices under the log-Cholesky, log-Euclidean, Frobenius, or power-Frobenius metrics, all of which are Hadamard spaces. Condition \textit{(A6)} is also satisfied in these settings. For distributional-valued response in Wasserstein space, let $F_{X_j},$ $F_{\oplus j},$ and $F_{\nu}$ be distribution functions, and $F_{X_j}^{-1},$ $F^{-1}_{\oplus j},$ and $F^{-1}_{\nu}$ their corresponding quantile functions of $g_j(x_j),$ $\mathbb{E}_{\oplus} g_j(X_j),$ and $\nu \in \M.$ Then the transport map is represented as a quantile function $T_j(X_j)(\nu) = F_{X_j}^{-1}\circ F_{\oplus j} \circ F^{-1}_{\nu},$ and its \F mean $F_{\oplus j}^{-1}\circ F_{\oplus j} \circ F^{-1}_{\nu} = F^{-1}_{\nu},$ so \textit{(A6)} is satisfied. For SPD matrices with the log-Cholesky metric, the Cholesky decomposition yields a vector representation in Euclidean space, which directly implies that Condition \textit{(A6)} holds. The same argument extends to other metrics on SPD matrices, including the Frobenius, and power-Frobenius metrics. The inequality in Condition \textit{(A6)} serves as a regularity condition analogous to \textit{(A3)}, ensuring the consistency of M-estimators and the uniqueness of the population \F mean.

\subsection{Proof of Theorem 1}

We need to show that for any $\nu \in \M$ and $x_j \in \mathcal{X}_j,$ if the bandwidth $h \to 0$ and $nh \to \infty,$ it holds that \begin{align*}
    d(T_j(x_j)(\nu), \hat{T}_j(x_j)(\nu) ) = o_p(1),
\end{align*} where $\hat{T}_j(\nu)$ is the estimates for the proposed transport backfitting algorithm.

For fixed $\nu \in \M$ and $x_j \in \mathcal{X}_j$ for fixed $j,$ we have \begin{align}
    d\left(T_j(x_j)(\nu), \hat{T}_j(x_j)(\nu)\right) &\le d\left(T_j(x_j)(\nu), \hat{T}^L_j(x_j)(\nu) \right) + d\left(\hat{T}^L_j(x_j)(\nu),  \hat{T}_j(x_j)(\nu) \right) 
    \label{eq:supp:target}
\end{align} By Lemma 1 and the proof Theorem 3 in the Supplement of \cite{petersen2019frechet}, under conditions \textit{(A1)-(A3)}, we observe the bias term \begin{align}
    d\left(T_j(x_j)(\nu), {T}^L_j(x_j)(\nu) \right) = o(1),
    \label{eq:supp:bias}
\end{align} as the bandwidth $h \to 0.$
Also, by Lemma 2 in the Supplement of  \cite{petersen2019frechet}, under conditions \textit{(A1)} and \textit{(A3)}, we observe the stochastic term \begin{align}
    d\left(T^L_j(x_j)(\nu), \hat{T}^L_j(x_j)(\nu) \right) = o_p(1)
    \label{eq:supp:stoc}
\end{align} as $h \to 0$ and $nh \to \infty.$ Combining \eqref{eq:supp:bias} and \eqref{eq:supp:stoc}, we have the first term of \eqref{eq:supp:target} \begin{align}
     d\left({T}_j(x_j)(\nu), \hat{T}^L_j(x_j)(\nu) \right) = o_p(1)
     \label{eq:supp:loss}
\end{align}

Thus, it is enough to show that the second term of \eqref{eq:supp:target} \begin{align}
    d\left(\hat{T}^L_j(x_j)(\nu),  \hat{T}_j(x_j)(\nu) \right) = o_p(1).
    \label{supp:eq:intermediate}
\end{align} By the transport backfitting algorithm we can represent, \begin{align*}
    \hat{T}_j(x_j)(\nu) = T_{\hat{\mathbb{E}}_{\oplus}\hat{g}_{j}(X_j), \hat{g}_{j}(x_j) }(\nu),
\end{align*}  where \begin{align*}
    \hat{g}_{j}(x_j) &= \argmin_{\omega \in \M } \frac{1}{n}\sum_{i=1}^{n} \hat{w}_i(x_j, h) d^2(P_{\oplus i, j}(\nu), \omega),\\
    \hat{\mathbb{E}}_{\oplus}\hat{g}_{j}(X_j) &= \argmin_{\omega \in \M} \frac{1}{n}\sum_{i=1}^{n} d^2(\hat{g}_{j}(X_{i,j}), \omega).
\end{align*} 
By condition \textit{(A6)} and the property of transport map, we have $\hat{T}^L_j(x_j)(\nu) = T_{\nu, \hat{T}^L_j(x_j)(\nu)}(\nu) = T_{\mathbb{E}_{\oplus} T_j(X_j)(\nu), \hat{T}^L_j(x_j)(\nu)}(\nu).$ The equation \eqref{supp:eq:intermediate} is \begin{align*}
    d\left(\hat{T}^L_j(x_j)(\nu),  \hat{T}_j(x_j)(\nu) \right) &= d\left( T_{\mathbb{E}_{\oplus} T_j(X_j)(\nu), \hat{T}^L_j(x_j)(\nu)}(\nu),~ T_{\hat{\mathbb{E}}_{\oplus}\hat{g}_{j}(X_j), \hat{g}_{j}(x_j) }(\nu)  \right) \\
    &\le C \left\{ d\left(\mathbb{E}_{\oplus} T_j(X_j)(\nu),\hat{\mathbb{E}}_{\oplus}\hat{g}_{j}(X_j)\right) + d \left(\hat{T}^L_j(x_j)(\nu), \hat{g}_{j}(x_j) \right) \right\},
\end{align*} for some constant $C >0,$ and the last inequality came from condition \textit{(A5)}.

\vspace{3mm}
\noindent \textbf{Lemma 1} \textit{Under (A1), (A3), (A4), and (A5), if the bandwidth $h \to 0$ and $nh \to \infty$ as $n \to \infty$, it holds that
\begin{align*}
     d \left(\hat{T}^L_j(x_j)(\nu), \hat{g}_{j}(x_j) \right) = o_{P}(1),
\end{align*} for any $x_j \in \mathcal{X}_j,$ and $\nu \in \M.$} 
\vspace{3mm}

\vspace{3mm}
\noindent \textbf{Lemma 2} \textit{Under (A1)-(A6), if the bandwidth $h \to 0$ and $nh \to \infty$ as $n \to \infty$, it holds that
\begin{align*}
     d\left(\mathbb{E}_{\oplus} T_j(X_j)(\nu),\hat{\mathbb{E}}_{\oplus}\hat{g}_{j}(X_j)\right) = o_{P}(1),
\end{align*} for any $x_j \in \mathcal{X}_j,$ and $\nu \in \M.$} 
\vspace{5mm}

Combining Lemma 1 and Lemma 2, we have $d \left(\hat{T}^L_j(x_j)(\nu), \hat{g}_{j}(x_j) \right) = o_p(1)$ and $d\left(\mathbb{E}_{\oplus} T_j(X_j)(\nu),\hat{\mathbb{E}}_{\oplus}\hat{g}_{j}(X_j)\right) = o_p(1),$ leading to the desired results in \eqref{supp:eq:intermediate},
\begin{align*}
    d\left(\hat{T}^L_j(x_j)(\nu),  \hat{T}_j(x_j)(\nu) \right) = o_p(1).
\end{align*}

Then, the equation \eqref{eq:supp:loss} and \eqref{supp:eq:intermediate} verifies our goal of \eqref{eq:supp:target}, \begin{align*}
    d\left(T_j(x_j)(\nu), \hat{T}_j(x_j)(\nu)\right) = o_p(1).
\end{align*}

\subsection{Proof of Lemmas}

\subsubsection{Proof of Lemma 1}

By Corollary 3.2.3 in \cite{van2023weak}, for the proof of the consistency of M-estimators $d \left(\hat{T}^L_j(x_j)(\nu), \hat{g}_{j}(x_j) \right) = o_p(1),$ it is sufficient to show that the uniform consistency of loss functions, \begin{align*}
    \sup_{\omega \in \M} \bigg\lvert \frac{1}{n} \sum_{i=1}^n \hat{w}_i(x_j, h) d^{2}\left(Z_{i,j}(X_{i,j})(\nu), \omega\right) -  \frac{1}{n}\sum_{i=1}^{n} \hat{w}_i(x_j, h) d^2(P_{\oplus i, j}(\nu), \omega)\bigg\rvert = o_p(1).
\end{align*} Combining Theorem  1.3.6 and 1.5.4 of \cite{van2023weak}, we need to show that \begin{enumerate}
    \item[(1)] (Pointwise consistency) $\frac{1}{n} \sum_{i=1}^n \hat{w}_i(x_j, h) d^{2}\left(Z_{i,j}(X_{i,j})(\nu), \omega\right) -  \frac{1}{n}\sum_{i=1}^{n} \hat{w}_i(x_j, h) d^2(P_{\oplus i, j}(\nu), \omega) = o_p(1)$ for all $\omega \in \M,$ and
    \item[(2)] (Asmptotically equicontinuous in probability) For all $\epsilon, \eta > 0,$ there exists $\delta > 0$ such that
    \begin{align*}
        \limsup_{n} P\left( \sup_{d(\omega_1, \omega_2) < \delta} \lvert A_n(\omega_1) - A_n(\omega_2) \rvert > \epsilon \right) < \eta,
    \end{align*} where $A_n(\omega) = \frac{1}{n} \sum_{i=1}^n \hat{w}_i(x_j, h) d^{2}\left(Z_{i,j}(X_{i,j})(\nu), \omega\right) -  \frac{1}{n}\sum_{i=1}^{n} \hat{w}_i(x_j, h) d^2(P_{\oplus i, j}(\nu), \omega).$
\end{enumerate}

Begin with (1), \begin{align*}
    &\bigg\lvert \frac{1}{n} \sum_{i=1}^n \hat{w}_i(x_j, h) \left\{d^{2}\left(Z_{i,j}(X_{i,j})(\nu), \omega\right) - d^2(P_{\oplus i, j}(\nu), \omega) \right\}  \bigg\rvert \\
    &= \bigg\lvert \frac{1}{n} \sum_{i=1}^n \hat{w}_i(x_j, h) \bigg\{d^{2}\left( \left[ \varepsilon_{i,n} \circ T_j (X_{i,j}) \right] (\nu), \omega\right) \\
    &\quad\quad\quad- d^2([T^{-1}_{j+1}(X_{i,j+1}) \circ \cdots \circ T^{-1}_{p}(X_{i,p}) \circ \varepsilon_{i,n} \circ  T_{p}(X_{i,p}) \circ \cdots \circ T_{j+1}(X_{i,j+1}) \circ T_{j}(X_{i,j})] (\nu), \omega) \bigg\}  \bigg\rvert \\
    &\le 2 \text{diam}(\M) \\
    &\times  \frac{1}{n} \sum_{i=1}^n \hat{w}_i(x_j,h) d\left( \left[ \varepsilon_{i,n} \circ T_j (X_{i,j}) \right] (\nu),[T^{-1}_{j+1}(X_{i,j+1}) \circ \cdots \circ T^{-1}_{p}(X_{i,p}) \circ \varepsilon_{i,n} \circ  T_{p}(X_{i,p}) \circ \cdots \circ T_{j}(X_{i,j})] (\nu) \right),
\end{align*} where the last inequality is from $d^2(a, \omega) - d^2(b,w) \le \lvert d(a,w) - d(b,w)\rvert (d(a,w) + d(b,w) \rvert \le 2 \text{diam}(\M) d(a,b),$ for any $a, b \in \M$, and fixed $\omega \in M.$

\vspace{3mm}
\noindent \textbf{Lemma 3} \textit{Under conditions (A4) and (A5), it holds that
\begin{align*}
    \frac{1}{n}\sum_{i=1}^{n} d \left( [T \circ \varepsilon_{i,n}](\xi), [\varepsilon_{i,n} \circ T]  (\xi)\right) = o_{p}(1),
\end{align*} for any $T \in \mathcal{T},$ and $\xi \in \M.$} 
\vspace{5mm}

By Lemma 3 with $T = T^{-1}_{j+1}(X_{i,j+1}) \circ \cdots \circ T^{-1}_{p}(X_{i,p}) \in \mathcal{T}$ and $\xi = [ T_{p}(X_{i,p}) \circ \cdots \circ T_{j}(X_{i,j})] (\nu) \in \M,$ we observe that \begin{align*}
     \frac{1}{n} \sum_{i=1}^n d\left( \left[ \varepsilon_{i,n} \circ T_j (X_{i,j}) \right] (\nu),[T^{-1}_{j+1}(X_{i,j+1}) \circ \cdots \circ T^{-1}_{p}(X_{i,p}) \circ \varepsilon_{i,n} \circ  T_{p}(X_{i,p}) \circ \cdots \circ T_{j}(X_{i,j})] (\nu) \right) = o_p(1).
\end{align*}
Also, by Lemma 2 in Supplement of \cite{petersen2019frechet}, $\frac{1}{n} \sum_{i=1}^n  \hat{w}_i(x_j,h)  = O_p(1),$ with $\hat{w}_i(x,h) > 0$ for all $x.$ Since $\M$ is bounded, we have \begin{align*}
    \bigg\lvert \frac{1}{n} \sum_{i=1}^n \hat{w}(x_j, h) \left\{d^{2}\left(Z_{i,j}(X_{i,j})(\nu), \omega\right) - d^2(P_{\oplus i, j}(\nu), \omega) \right\}  \bigg\rvert = o_p(1),
\end{align*} which is the pointwise consistency in (1).

Moving on to (2), \begin{align*}
    \lvert A_n(\omega_1) - A_n(\omega_2)  \rvert &= \frac{1}{n}\sum_{i=1}^{n} \hat{w}_i(x_j, h) \left\{d^2(Z_{i,j}(X_{i,j})(\nu), \omega_1) -d^2(Z_{i,j}(X_{i,j})(\nu), \omega_2) \right\} \\
    &\quad\quad + \frac{1}{n}\sum_{i=1}^{n} \hat{w}_i(x_j, h) \left\{d^2(P_{\oplus i, j}(\nu), \omega_1) -d^2(P_{\oplus i, j}(\nu), \omega_2) \right\} \\
    &\le 4\text{diam}(\M) d(\omega_1, \omega_2) \frac{1}{n} \sum_{i=1}^{n} \lvert \hat{w}_i(x_j,h) \rvert.
\end{align*} Then, $\frac{1}{n} \sum_{i=1}^n \lvert \hat{w}_i(x_j,h) \rvert = O_p(1)$ leads to $ \lvert A_n(\omega_1) - A_n(\omega_2)  \rvert = O_{p}(d(\omega_1, \omega_2)),$ which verifies (2).

\subsubsection{Proof of Lemma 2}
We observe \begin{align*}
    d\left(\mathbb{E}_{\oplus} T_j(X_j)(\nu),\hat{\mathbb{E}}_{\oplus}\hat{g}_{j}(X_j)\right) \le d\left(\mathbb{E}_{\oplus} T_j(X_j)(\nu),\hat{\mathbb{E}}_{\oplus}T_j(X_j)(\nu)\right) + d\left(\hat{\mathbb{E}}_{\oplus} T_j(X_j)(\nu),\hat{\mathbb{E}}_{\oplus}\hat{g}_{j}(X_j)\right).
\end{align*}
By \cite{dubey2019frechet}, under condition \textit{(A6),} the consistency of sample \F mean holds, i.e., \begin{align}
    d\left(\mathbb{E}_{\oplus} T_j(X_j)(\nu),\hat{\mathbb{E}}_{\oplus}T_j(X_j)(\nu)\right) = o_p(1).
    \label{eq:supp:lemma2_pt1}
\end{align}
Following proof of Lemma 1, for the proof of $d\left(\hat{\mathbb{E}}_{\oplus} T_j(X_j)(\nu),\hat{\mathbb{E}}_{\oplus}\hat{g}_{j}(X_j)\right) = o_p(1),$ it is sufficient to show that \begin{align*}
    \sup_{\omega \in M} \bigg\lvert \frac{1}{n} \sum_{i=1}^n d^2(\hat{g}_j(X_{i,j}), \omega) -  \frac{1}{n} \sum_{i=1}^n d^2(T_j(X_{i,j}), \omega)  \bigg\rvert = o_p(1).
\end{align*}
We observe \begin{align*}
    &\sup_{\omega \in M} \bigg\lvert \frac{1}{n} \sum_{i=1}^n d^2(\hat{g}_j(X_{i,j}), \omega) -  \frac{1}{n} \sum_{i=1}^n d^2(T_j(X_{i,j}), \omega)  \bigg\rvert \le 2 \text{diam}(\M) \frac{1}{n} \sum_{i=1}^n d(\hat{g}_j(X_{i,j}),T_j(X_{i,j})) \\
    &\quad\quad\quad\le 2\text{diam}(\M) \frac{1}{n} \sum_{i=1}^n \left\{d(\hat{g}_j(X_{i,j}),\hat{T}^L_j(X_{i,j}))  + d(\hat{T}^L_j(X_{i,j}),T_j(X_{i,j})) \right\}.
\end{align*}
By Lemma 1, the first term $\frac{1}{n} \sum_{i=1}^n d(\hat{g}_j(X_{i,j}),\hat{T}^L_j(X_{i,j})) = o_p(1)$ and by equation \eqref{eq:supp:loss}, the second term $\frac{1}{n} \sum_{i=1}^n d(\hat{T}^L_j(X_{i,j}),T_j(X_{i,j})) = o_p(1).$ Thus, we have \begin{align}
    d\left(\hat{\mathbb{E}}_{\oplus} T_j(X_j)(\nu),\hat{\mathbb{E}}_{\oplus}\hat{g}_{j}(X_j)\right) = o_p(1).
    \label{eq:supp:lemma2_pt2}
\end{align}

Combining \eqref{eq:supp:lemma2_pt1} and \eqref{eq:supp:lemma2_pt2}, we verify \begin{align*}
    d\left(\mathbb{E}_{\oplus} T_j(X_j)(\nu),\hat{\mathbb{E}}_{\oplus}\hat{g}_{j}(X_j)\right) = o_p(1).
\end{align*}

\subsubsection{Proof of Lemma 3}
For fixed $T \in \mathcal{T},$ and $\xi \in \M,$ we observe
\begin{align*}
     d \left( [T \circ \varepsilon_{i,n}](\xi), [\varepsilon_{i,n} \circ T]  (\xi)\right) &\le   d \left( [T \circ \varepsilon_{i,n}](\xi), T (\xi)\right) +  d \left( T(\xi), [\varepsilon_{i,n}\circ T]  (\xi)\right) \\
     &\le C  d \left( \varepsilon_{i,n}(\xi), \xi \right) + d \left( T(\xi), [\varepsilon_{i,n}\circ T]  (\xi)\right),
\end{align*} for constant $C > 0.$ The first inequality is from triangle inequality, and second inequality is from condition \textit{(A5)}. 

Then, \begin{align*}
    \frac{1}{n}\sum_{i=1}^{n} d \left( [T \circ \varepsilon_{i,n}](\xi), [\varepsilon_{i,n} \circ T]  (\xi)\right) &\le C  \frac{1}{n}\sum_{i=1}^{n} d \left( \varepsilon_{i,n}(\xi), \xi \right) + \frac{1}{n}\sum_{i=1}^{n} d \left( T(\xi), [\varepsilon_{i,n}\circ T]  (\xi)\right) \\
    &\to C \mathbb{E} [d(\varepsilon_{1,n}(\xi), \xi)] + \mathbb{E}[d(\varepsilon_{1,n}(T(\xi)),T(\xi) )] \quad \text{by WLLN}.
\end{align*} By condition \textit{(A4)}, we have \begin{align*}
    \frac{1}{n}\sum_{i=1}^{n} d \left( [T \circ \varepsilon_{i,n}](\xi), [\varepsilon_{i,n} \circ T]  (\xi)\right) = o_{p}(1),
\end{align*} which is the desired results.

\section{Further details on Brain connectivity for fMRI data analysis}

\subsection{Data Preprocessing}
Brain signal analysis at the subject level relies  on time series of Blood Oxygen Level Dependent (BOLD) signals, into a set of regions of interest (ROI). The coherence between pairwise ROIs is usually measured by Pearson correlation coefficients of the fMRI time series, leading to $l \times l$ correlation matrix for $l$ ROIs. Alzheimer's disease has been found to be associated with anomalies in the functional integration of ROIs \citep{badhwar2017resting, perovnik2023functional}. The list of $l = 11$ ROIs used in the analysis is provided in Table \ref{supp:tbl1}. Preprocessing was conducted in MATLAB using the Statistical Parametric Mapping (SPM12, \url{www.fil.ion.ucl.ac.uk/spm}) and the rs-fMRI Data Analysis Toolkit V1.8 (REST1.8, \url{https://rfmri.org/REST}); further details are available in \cite{zhou2022network}.

\begin{table}[h!]
\caption{Brain region of interest (ROI) used in the analysis along with its ROI label \citep{andrews2010functional}.}
\label{supp:tbl1}
\centering
\begin{tabular}{l|l}
\toprule
Region Full Name & ROI Label \\
\midrule
Anterior medial prefrontal cortex & MPFC \\
\addlinespace
Posterior cingulate cortex & PCC \\
\addlinespace
Dorsal medial prefrontal cortex & dMFPC \\
\addlinespace
Temporal parietal junction & TPJ \\
\addlinespace
Lateral temporal cortex & LTC \\
\addlinespace
Temporal pole & TempP \\
\addlinespace
Ventral medial prefrontal cortex & vMFPC \\
\addlinespace
Posterior inferior parietal lobule & pIPL \\
\addlinespace
Retrosplenial cortex & Rsp \\
\addlinespace
Parahippocampal cortex & PHC \\
\addlinespace
Hippocampal formation & HF+ \\
\bottomrule
\end{tabular}
\end{table}

{
\subsection{Additional Simulations}

In this subsection, we provide out-of-sample prediction performance comparison results for a stratified analysis where we use stage-specific \F means. Subjects were grouped into three diagnostic stages:
\begin{itemize}
    \item \textbf{Group 1} (358 subjects): Cognitively Normal (CN), Subjective Memory Complaint (SMC),
    \item \textbf{Group 2} (499 subjects): Early Mild, Mild, and Late Mild Cognitive Impairment (EMCI, MCI, LMCI),
    \item \textbf{Group 3} (72 subjects): Alzheimer’s Disease (AD).
\end{itemize}

We evaluate out-of-sample prediction performance using 5-fold cross-validation and  partition the index set ${1,\ldots,n}$ into disjoint folds $\mathcal{I}_q$ $(q=1,\ldots,5)$ with $n_q = |\mathcal{I}_q|$. For each fold, we define the refined Mean Prediction Error Reduction (rMPER) as
\begin{align*}
    \text{rMPER} &=  \frac{1}{5}\sum_{q=1}^{5} \text{rMPER}_{q}, \\
    \text{rMPER}_{q} &= \frac{1}{n_{q}} \sum_{i \in \mathcal{I}_q} \left\{ d(\hat{\mu}_{i,\oplus}^{(-q)}, Y_i) - d(\hat{Y}^{(-q)}_i, Y_i) \right\},
\end{align*}  where
$\hat{Y}_i^{(-q)}$ is the predicted response from the regression model trained on the remaining folds, $\hat{\mu}_{i,\oplus}^{(-q)}$ is the disease-profile \F mean corresponding to the $i$th subject's group (Group 1, 2, or 3) in the training set of fold $q,$ and $d$ is the log-Cholesky metric. rMPER measures the reduction in prediction error relative to the disease-profile baseline; positive rMPER values indicate improved prediction relative to the disease-profile baseline. 
We compare our method against global \F regression (GF) \citep{petersen2019frechet}. Table \ref{supp:tbl2} reports 
the average of rMPER $(\times 10^{-4})$ obtained for  $B=200$ Monte Carlo simulations  for ADOPT and global \F regression. ADOPT achieves a larger rMPER, indicating better out-of-sample predictive performance than global \F regression.

\begin{table}[h!]
\caption{Mean Prediction Error Reduction (MPER $\times 10^{-4}$) with standard errors in parentheses, based on ADNI brain correlation regression analysis with log-Cholesky metric. Higher MPER values indicate better out-of-sample prediction. We compare two methods: additive optimal transport regression (ADOPT), and global \F regression \citep{petersen2019frechet} (GF)}
\label{supp:tbl2}
\centering
\begin{tabular}{ccc}
\toprule
&\textbf{ADOPT} & \textbf{GF} \\
\midrule
MPER & \textbf{3.020 (0.0526)}  & 2.108 (0.0518) \\
\bottomrule
\end{tabular}
\end{table}

To check the robustness of the ADOPT model, we repeat the following procedure 1000 times: Each time 1) fitting the ADOPT model on a random 80\% subsample and 2) identifying the strongest signal among 55 ROI-to-ROI correlations. The PIPL–MPFC connection appears the strongest signal in 545 of 1000 runs, consistent with the full-data result.

}

{\section{Computational Complexity Analysis}
The computational complexity of ADOPT depends on the underlying metric space $(\M, d).$ Let $n$ be the number of observations, and $p$ be the dimension of predictors. For each cycle of transport backfitting iteration, roughly speaking, the overall complexity is $O(n^2p + np^2),$ which simplifies to $O(n^2p)$ in the common setting where $p << n.$ For specific object spaces $(\M, d)$:
\begin{itemize}
    \item Univariate distributions with 2-Wasserstein metric: When probability distributions are represented by quantile functions with $q$ discretization points, the complexity becomes $O(n^2pq).$
    \item Symmetric positive definite (SPD) matrices with Log-Cholesky metric: For $m \times m$ SPD matrices, operations scale with the number of lower-triangular components. In this case, the complexity is $O(n^2pm^2).$
\end{itemize}
}

\end{document}